\documentclass[10pt,
footinbib,
bibnotes,
amsmath,amssymb,
aps,
prb,
twocolumn,
]{revtex4-1}
\usepackage{natbib}
\usepackage{ifthen}
\usepackage{color}
\usepackage{multirow}
\usepackage{xcolor}
\usepackage{graphicx}
\usepackage{epsfig}
\usepackage[hidelinks]{hyperref}
\hypersetup{
	pdfstartview=FitH,
	colorlinks=true,
	linkcolor=blue,
	citecolor=blue,
	urlcolor= blue
}
\usepackage[normalem]{ulem}
\usepackage{latexsym}
\usepackage{amsmath}
\usepackage{amssymb}
\usepackage{bm}
\usepackage{wasysym}
\usepackage{gensymb}
\usepackage{overpic}
\usepackage{latexsym}
\usepackage{amsmath}
\usepackage{amssymb}
\usepackage{bm}
\usepackage{wasysym}
\usepackage{float}
\usepackage{braket}
\usepackage{svg}

\begin{document}
	\title{Spectral Properties of Disordered Interacting Non-Hermitian Systems}
\author{Soumi Ghosh}
\email{soumi.ghosh@icts.res.in } 
\address{International Centre for Theoretical Sciences, Tata Institute of Fundamental Research, Bengaluru -- 560089, India}
\author{Sparsh Gupta}
\email{sparsh.gupta@icts.res.in  }
\address{International Centre for Theoretical Sciences, Tata Institute of Fundamental Research, Bengaluru -- 560089, India}
\author{Manas Kulkarni}
\email{manas.kulkarni@icts.res.in}
\address{International Centre for Theoretical Sciences, Tata Institute of Fundamental Research, Bengaluru -- 560089, India}
\date{\today}

\begin{abstract}
Non-hermitian systems have gained a lot of interest in recent years. However, notions of chaos and localization in such systems have not reached the same level of maturity as in the Hermitian systems. Here, we consider non-hermitian interacting disordered Hamiltonians and attempt to analyze their chaotic behavior or lack of it through the lens of the recently introduced non-hermitian analog of the spectral form factor and the complex spacing ratio. We consider three widely relevant non-hermitian models which are unique in their ways and serve as excellent platforms for such investigations. Two of the models considered are short-ranged and have different symmetries. The third model is long-ranged, whose hermitian counterpart has itself become a subject of growing interest. All these models exhibit a deep connection with the non-hermitian Random Matrix Theory of corresponding symmetry classes at relatively weak disorder. At relatively strong disorder, the models show the absence of complex eigenvalue correlation, thereby, corresponding to Poisson statistics. Our thorough analysis is expected to play a crucial role in understanding disordered open quantum systems in general.
\end{abstract}
\maketitle
\section{Introduction}
The emergence of quantum chaos and thermalization in closed interacting quantum systems has attracted a lot of interest in recent years~\cite{Deutsch.1991,Chaos_Srednicki.1994,Rigol.2008,QChaos_DAlessio.2016} owing to the occurrence of thermalization despite unitary evolution. On the other hand, the absence of thermalization in isolated disordered interacting systems dubbed many-body localization (MBL)~\cite{review_Nandkishore.2015,review_Alet.2018,review_Abanin.2019} has claimed huge attention due to its fascinating properties like logarithmic increase in entanglement entropy~\cite{EEgrowth_Bardarson.2012,EEgrowth_Serbyn.2013}, area law entanglement entropy for eigenstates with non-zero energy density away from the ground states~\cite{Arealaw_Bauer.2013,Zindaric.2008,Conservation_Serbyn.2013}, emergent integrability~\cite{lbit-huse.2014,Conservation_Serbyn.2013}, absence of energy level repulsion~\cite{Oganesyan.2007,Pal.2010}, to name a few. In realistic systems, the effects of external environment are inevitable which makes understanding the fate of localization in open quantum systems very crucial~\cite{MBL_OQS.2017,MBL_OQS.2018,InteractionDisorder_OQS.2018}. Moreover, attempts to engineer controlled environments~\cite{murch2012cavity, kimchi2016stabilizing} have been very successful leading to non-trivial phenomena in open quantum systems, which are often described by the Lindblad quantum master equations~\cite{carmichael1999statistical,breuer2002theory}. An alternate direction of investigation has been to directly study effective non-hermitian Hamiltonians, which often arise when certain conditions are imposed on open quantum systems~\cite{Daley.2014,atomics_tony.2014}.

Non-hermitian random matrices have been thoroughly investigated~\cite{Ginibre.1965,Sommers.1988,Sommers.1991} and  non-hermitian physics has attracted a lot of interest recently in the context of dissipative systems~\cite{Grobe.1988,Chalker.1997,Muller_OQS.2012,Can_dissipative.2019,Denisov_Lindblad.2019,Lucas_Lindblad.2020}, non-hermitian optics~\cite{PTsymmetry_Makris.2008,PTsymmetry_Klaiman.2008,PTsymmetry_Ruter.2010,PTsymmetry_review.2016,PTsymmetry_Feng.2017,expt_review.2018,expt_review.2020,tzortzakakis2020non}, topological phases in open systems~\cite{NH_topology1.2015,NH_topology5.2016,NH_topology7.2017,NH_topology2.2018,NH_topology3.2018,NH_topology4.2018,NH_topology6.2018,NH_topology9.2018,NH_topology10.2018,NH_topology11.2018,NH_topology8.2019,NH_topology12.2019,NH_topology14.2020,NH_topology13.2021,ExptSkineffect.2021,tang2020topological,kawabata2022topology,tang2021}, etc.
Very recently a many body localized phase has been investigated in interacting non-hermitian systems with disorder~\cite{Panda_NHMBL.2020,NHMBL_hamazaki.2019} and quasiperiodicity~\cite{NHMBL-QP_Zhai.2020}. Understanding different aspects of localization and chaos in closed quantum systems have reached a certain level of maturity. However, open quantum systems / non-hermitian systems are far from being thoroughly understood.
In this context, effort is needed in both identifying suitable platforms of open quantum systems and appropriate diagnostics to probe them.

Spectral properties have been used quite extensively to understand quantum chaos in ergodic quantum systems owing to the conjecture which states that chaotic quantum systems have spectral correlations similar to random matrices belonging to the corresponding symmetry classes~\cite{BGS.1984}. On the other hand, the uncorrelated real energy levels having Poisson statistics correspond to integrable systems according to the Berry-Tabor conjecture~\cite{BerryTabor.1977}. Various spectral properties have been used to understand the ergodic-MBL transition in interacting disordered systems~\cite{Oganesyan.2007}. Because of the growing interest in non-hermitian physics, some of these properties have been generalized for non-hermitian Hamiltonians with complex eigenvalues to understand chaos or lack thereof~\cite{NHMBL_hamazaki.2019,ProseLS.2019,cspacing_lucas.2020}. Specifically, level spacing statistics and generalized complex spacing ratio have been calculated for open systems with the Lindbladian approach~\cite{wang2020,yusipov2022,garcia2022, hamazaki2022lindbladian}, non-hermitian interacting disordered~\cite{NHMBL_hamazaki.2019,csr_syk2022,suthar2022non,Xiao.2022} and quasiperiodic~\cite{NHMBL-QP_Zhai.2020} systems. The complex spacing ratio has also been calculated for non-hermitian Dirac operators~\cite{Qcd2021}, dissipative quantum circuits~\cite{prosen2021circuits}, and non-interacting, non-hermitian disordered models in higher dimensions~\cite{Luo2021,Yan2022}. While both these quantities capture correlations between nearest-neighbor energy levels, another quantity of interest, called the dissipative spectral form factor, has been recently proposed and calculated for non-hermitian random matrices and several toy models~\cite{dsff_Li.2021,shivam2022many}. 
This quantity captures long-range correlations among the energy levels in the complex plane.
 \begin{figure}[t]
    \includegraphics[width=0.4\textwidth]{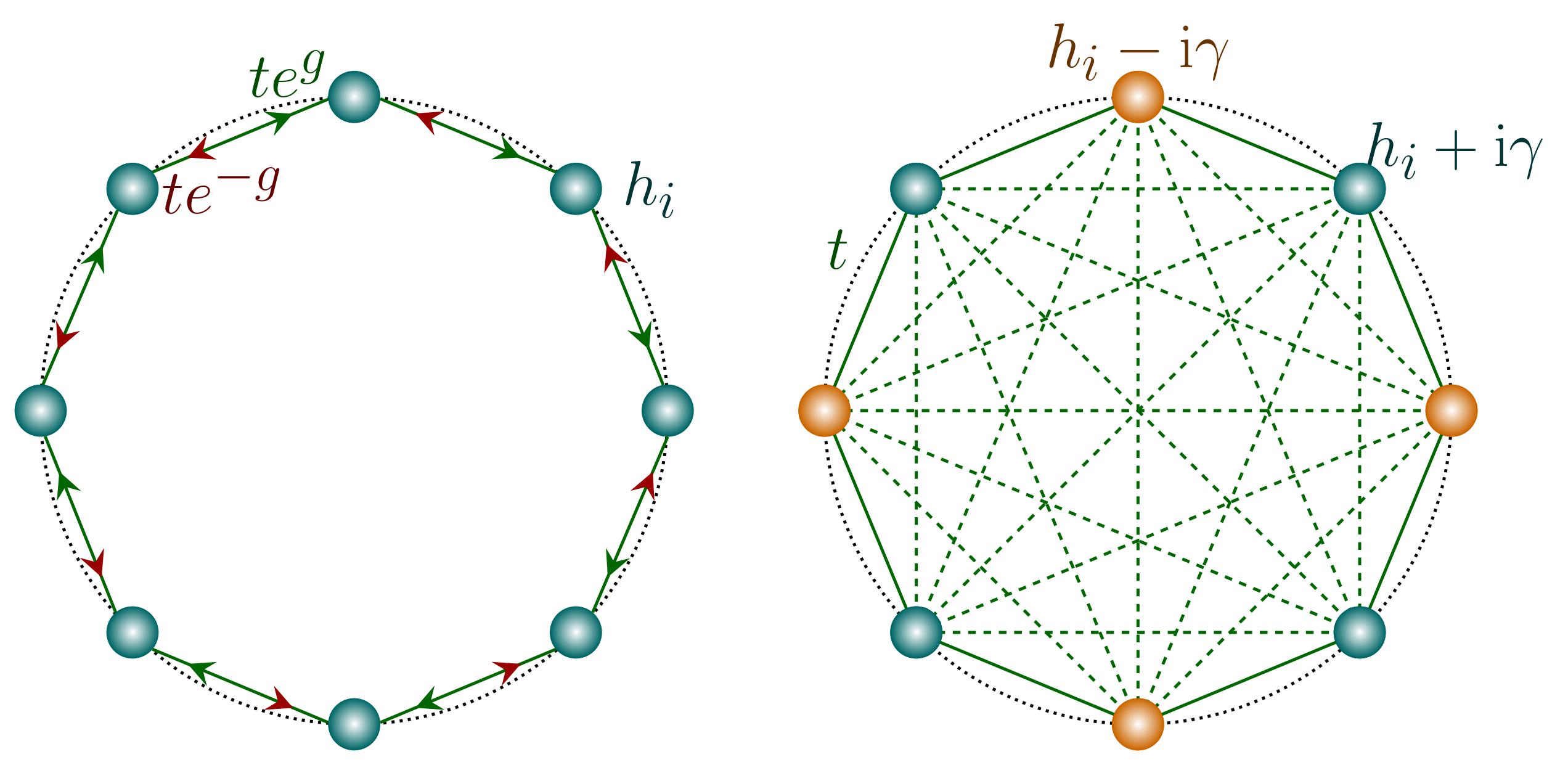}
    \caption{Schematic diagram showing the three models for representative system size. The diagram on the left depicts model-I (Eq.~\ref{eq:model1}), where each lattice site represented by a circle is associated with random onsite potential. The oppositely directed arrows of different colors represent the non-reciprocal hopping parameter. The diagram in the right represents both model-II (Eq.~\ref{eq:model2}) and model-III (Eq.~\ref{eq:model3}). Alternate teal-blue and orange circles have imaginary contributions $+\mathrm{i}\gamma$ and $-\mathrm{i}\gamma$ respectively with the real parts given by random onsite potentials. For model-II, the dashed connections do not exist as it has only nearest-neighbor hopping. For model-III, all the solid and dashed connections are non-zero with the underlying hopping parameters depending on the distance.}
    \label{fig:model}
\end{figure}

Here, we investigate the dissipative spectral form factor (DSFF) and complex spacing ratio (CSR) for disordered interacting non-hermitian models which are of experimental interest~\cite{atomics_tony.2014,Expt_MBL2d.2016,expt_review.2018,expt_review.2020,ExptSkineffect.2021}. These models have an inhomogeneous (non-uniform density of states) spectrum in the complex plane with outliers at the edge of the spectrum unlike the uniform homogeneous spectrum of non-hermitian random matrices. We specifically consider three different models (see Fig.~\ref{fig:model}). Two of these are short-range (and have different symmetries), while the third one is a long-range model. While many body localization was initially proposed for closed quantum systems with short-range Hamiltonians, soon the effects of long-range interactions and long-range hopping were considered~\cite{Nandkishore_longrange.2017,Nag_longrange.2019,Roy_longrange.2019,Prasad_longrange.2021} owing to the huge relevance of long-range Hamiltonians in experiments~\cite{MBLlr_expt.2016,timecrystal_expt.2017a,timecrystal_expt.2017b,timecrystal_expt.2018}. In Hermitian long-range systems, it has been established that long-range hopping induces delocalization while localization persists in presence of long-range interactions~\cite{Nag_longrange.2019,Roy_longrange.2019}. Hence, it is natural to ask what is the effect of long-range hopping in non-hermitian systems using the third model as our platform.

We show that, despite the non-uniform density (clustering of eigenvalues) in the complex plane, the DSFF in these non-hermitian systems possess the universal dip-ramp-plateau feature of the non-hermitian random matrices in the quantum chaotic regime. In fact, the ramp has a power-law variation with time, which is a characteristic of the Ginibre random matrices. Similarly, the CSR for these models shows agreement with RMT predictions in this regime. On the other hand, for large disorder strength, when the system possesses randomly distributed eigenvalues, the DSFF loses the characteristic dip-ramp-plateau structure. Instead, after an initial dip and fluctuation at small time scales $|\tau|$, it saturates to $1$ indicating the lack of correlation among the eigenvalues in the large disorder regime. The CSR for such disorder strengths shows Poisson statistics.

The rest of the paper is as follows. In Sec.~\ref{sec:models}, we describe the non-hermitian models that we consider. In Sec.~\ref{sec:results}, we define the two quantities of interest, namely the dissipative spectral form factor and the complex spacing ratio. We discuss their behavior in different parameter regimes of our models and compare these with those of the non-hermitian random matrices in the weak disorder regime and Poisson statistics in the presence of strong disorder. Finally, in Sec.~\ref{sec:conclusion}, we summarize our results along with an outlook. Certain details of the calculations are relegated to the appendices.
\section{Models}\label{sec:models}
In this paper, we consider three different non-hermitian models in one dimension (see Fig.~\ref{fig:model}). These are interacting systems of hard-core Bosons with random disordered potential.
The non-hermitian feature is achieved either by implementing a non-reciprocal hopping (model-I) or by ensuring the onsite potentials are complex (model-II and III).

Model-I--- In this model, the particles hop through the one dimensional disordered lattice with a non-reciprocal nearest-neighbor hopping parameter where the non-reciprocity is induced by $g\in \mathbb{R}$. This is a generalization of the well-known non-interacting Hatano-Nelson model~\cite{Hatano.1996,Hatano.1997,Hatano.1998,Hatano.1998_2}. The Hamiltonian is given by [see Fig.~\ref{fig:model}(left)]
\begin{equation}\label{eq:model1}
H=\sum\limits_{i=1}^L \left[-t\left( e^{g}\hat{c}^\dagger_{i}\hat{c}_{i+1}^{\phantom{\dagger}}+e^{-g}\hat{c}^\dagger_{i+1}\hat{c}_{i}^{\phantom{\dagger}}\right)+h_i\hat{n}_{i}^{}+V\hat{n}_{i}\hat{n}_{i+1}\right].
\end{equation}
Here, $\hat{c}_i^\dagger$ is the creation operator corresponding to the creation of hard-core bosons at site $i$ and $\hat{n}_i=\hat{c}_i^\dagger \hat{c}_i$ counts the number of particles at site $i$. Hence, for our system of hard-core bosons, $\hat{n}_i$ can have any of the two possible values $1$ and $0$. Here, $t \in \mathbb{R}$ contributes to the hopping parameter (reciprocal part), and $V\in \mathbb{R}$ is the nearest-neighbor interaction.
The onsite potentials are completely real and randomly chosen from a uniform distribution $h_i\in\left[-h,h\right]$. This Hamiltonian preserves time-reversal symmetry and is known to have a complex-real transition of eigenvalues as well as an MBL transition as the disorder strength $h$ is tuned~\cite{Panda_NHMBL.2020,NHMBL_hamazaki.2019}.

Model-II--- This is a gain-loss disordered model where the real part of the onsite potential is random and and chosen from a uniform distribution $h_i\in\left[-h,h\right]$.
The imaginary parts of the onsite potential are given by the gain-loss parameter $\pm i \gamma$ with an alternating sign across the lattice. The Hamiltonian reads [see Fig.~\ref{fig:model}(right)]
\begin{eqnarray}\nonumber
H=\sum\limits_{i=1}^L \left[-t\left(\hat{c}^\dagger_{i}\hat{c}_{i+1}^{\phantom{\dagger}}+\hat{c}^\dagger_{i+1}\hat{c}_{i}^{\phantom{\dagger}}\right)\right.+&&\left.\left(h_i+\mathrm{i} \gamma (-1)^i\right)\hat{n}_{i}^{}\right.\\
&&\left.+V\hat{n}_{i}\hat{n}_{i+1}\right]\quad. \label{eq:model2}
\end{eqnarray}
This Hamiltonian breaks the time-reversal symmetry and does not possess any complex-real transition for the eigenvalues. However, an MBL transition has been predicted numerically for this model~\cite{NHMBL_hamazaki.2019,cspacing_lucas.2020}. This Hamiltonian has a transposition symmetry ($H=H^T$).

Model-III--- While the previous two models are short-ranged with only nearest-neighbor hopping and nearest-neighbor interactions, this is a long-ranged generalization of the model-II. Here, every site is coupled to all other sites in the lattice through hopping. However, the interaction remains nearest-neighbor only. The Hamiltonian in this case reads [see Fig.~\ref{fig:model}]
\begin{eqnarray}\nonumber
H=\sum\limits_{i,j=1}^{L}\dfrac{-t\phantom{|i|^\alpha}}{|i-j|^\alpha}\left(\hat{c}_i^\dagger c_j^{\phantom{\dagger}}+\hat{c}_j^\dagger \hat{c}_i^{\phantom{\dagger}}\right)+\sum\limits_{i=1}^L &&\left[\left(h_i+\mathrm{i} \gamma (-1)^i\right)\hat{n}_{i}^{}\right.\\
&&\left.+V\hat{n}_{i}\hat{n}_{i+1}\right]. \label{eq:model3}
\end{eqnarray}
The range of hopping is controlled by the parameter $\alpha$, where for a non-zero $\alpha$, the hopping strength decreases as the distance between the sites increases. In the extreme case of $\alpha=0$, each site is coupled to every other site with the same hopping strength $t$. On the other hand, the system with only nearest-neighbor hopping corresponds to the limit $\alpha\rightarrow\infty$. The Hamiltonian of this model has the same transposition symmetry as model-II (Eq.~\ref{eq:model2}).

All the three Hamiltonian commute with the total particle number operator $\hat{N}=\sum\limits_{i=1}^L \hat{n}_i$. Therefore, the Hamiltonian can be written in a block diagonal form in the particle number basis, where basis states having the same number of particles form a sector.
Hence, the Hamiltonian can be solved in different particle sectors independently. Here, we consider the half-filled sector for even system sizes $L$, in which case the dimension of the Hilbert space is given by $\mathcal{N}=\binom L {L/2}$. We exactly diagonalize the Hamiltonian and look for spectral properties for system sizes up to $L=18$. All the quantities are calculated taking disorder samples where the sample size are $4000, 200, 100 $ for $L=14, 16, 18$ respectively. We re-center the spectrum of each sample at $(0,0)$ in the complex plane and divide the energy eigenvalues by the maximum of the absolute values of eigenvalues in the spectrum for each sample. This ensures that the eigenvalue spectrum of each sample spreads over the same region in the complex plane. We choose the parameters as $t=1.0$, $V=2.0$, $g=0.1$, and $\gamma=0.1$. We compare our spectral statistics results with the corresponding random matrices of dimension $10000 \times 10000$ unless specified.
\section{Results}\label{sec:results}
In this section, we discuss our quantities of interest along with our findings on the three models (Eqns.~\ref{eq:model1},\ref{eq:model2},\ref{eq:model3}) discussed above.
\subsection{Dissipative Spectral Form Factor}
For a generic non-hermitian system (of dimension $\mathcal{N}\times\mathcal{N}$) having complex eigenvalues given by $z_n=x_n+\mathrm{i}y_n$, the density of states in the complex plane can be defined as
\begin{equation}\label{eq:dos} 
\rho(z)=\sum\limits_{n=1}^{\mathcal{N}} \delta^{(2)}(z-z_n)=\sum\limits_{n=1}^\mathcal{N}\delta(x-x_n)\delta(y-y_n)
\end{equation}
Similarly, the two-point correlation function for such a spectrum can be defined as \(\langle \rho(z_1)\rho(z_2+\omega) \rangle\) where $\omega$ is a complex variable \(\omega=\omega_x+\mathrm{i}\omega_y\) and \(\langle .\rangle\) denotes average over different ensembles. Both these quantities map to their well-known hermitian counterparts when the complex parts (\(y_n,\omega_y)\) do not exist. In the case of hermitian Hamiltonians, the spectral form factor (SFF) $K(t)$ is defined as the Fourier transform of the two-point correlation function in the time ($t$) domain. A similar quantity for the non-hermitian case thus requires a generalized time variable $\tau=t+\mathrm{i}s$ having both real and imaginary parts.
Thus, the non-hermitian analog of SFF, namely dissipative spectral form factor (DSFF) is defined as a function of \((t,s)\) or equivalently as a function of \((\tau,\tau^*)\) as follows~\cite{dsff_Li.2021}
\begin{equation}\label{eq:dsff}
K(\tau,\tau^*)=\dfrac{1}{\mathcal{N}}\left\langle \sum \limits_{n,m=1}^{\mathcal{N}}e^{-\mathrm{i}\vec{z}_{mn}.\vec{\tau}} \right\rangle
\end{equation}
Here \(\vec{z}_{mn}=(x_m-x_n,y_m-y_n)\) is the difference between the two eigenvalues \(z_m\) and \(z_n\). The dot product in the parenthesis represents the product \(\vec{z}_{mn}.\vec{\tau}=(x_m-x_n)t+(y_m-y_n)s\). The DSFF captures the long-range correlations among the real and the imaginary parts of the eigenvalues for all energy scales. Often it is useful to define the connected part of the DSFF $K_{c}(\tau,\tau^*)$ as
\begin{equation}
    K_c(\tau,\tau^*)=\dfrac{1}{\mathcal{N}}\left[\left\langle \sum \limits_{n,m=1}^{\mathcal{N}}e^{-\mathrm{i}\vec{z}_{mn}.\vec{\tau}} \right\rangle-\left|\left\langle\sum\limits_{m=1}^{\mathcal{N}} e^{-\mathrm{i}\vec{z}_m.\vec{\tau}}\right\rangle\right|^2\right]
\end{equation}

The DSFF for random ($\mathcal{N}\times \mathcal{N}$) non-hermitian Ginibre ensembles (GinUE, GinOE, GinSE) show the dip-ramp-plateau structure as a function of $|\tau|$ much like its hermitian counterpart, namely the SFF, for the hermitian random matrix ensembles~\cite{dsff_Li.2021}. For the case of the non-hermitian random matrices, although DSFF is defined for the complex spectra, the variation of DSFF is independent of $\phi=\mathrm{arg}(\tau)$ owing to the uniform and rotationally symmetric distribution of the energies $z_n$, or more precisely, due to the rotational symmetry of the distribution of $z_{mn}=z_m-z_n$~\cite{dsff_Li.2021}. Such rotational symmetry observed in random matrices is very unlikely to exist in the disordered Hamiltonians. Nonetheless, it turns out that there exists a deep connection between the Hamiltonians and the random matrices irrespective of the inhomogeneity and anisotropy of the underlying complex spectrum.\\
\begin{figure}[t]
\centering
\includegraphics[width=0.5\textwidth]{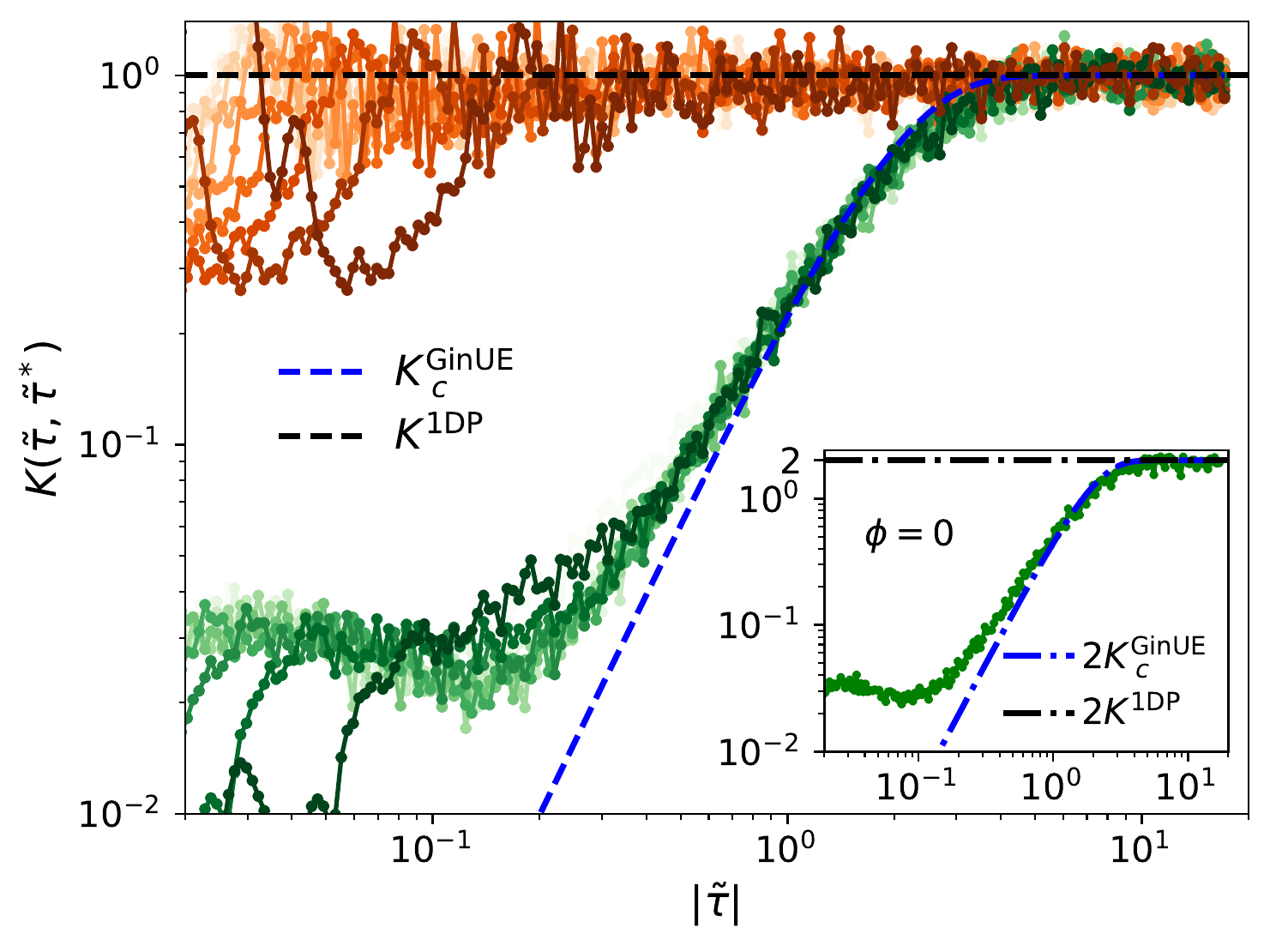}
\caption{[Model-I, Eq.~\ref{eq:model1}] DSFF for model-I as a function of rescaled time variable $|\tilde{\tau}|$ for different values of $\phi \in [\pi/20,9\pi/20]$ in steps of $\pi/20$ for a low disorder strength $h=2$ (Green) and a high disorder strength $14$ (Orange). The lowest value of $\phi$ in the given range corresponds to the lightest shade, while the highest value of $\phi$ corresponds to the darkest shade of the corresponding colors. The blue dashed line corresponds to the DSFF for GinUE ($K^{\mathrm{GinUE}}_c$), while the black line corresponds to that of uncorrelated random energy levels ($K^{\mathrm{1D P}}=1$).  We see evidence of a non-linear ramp in the DSFF for model-I at weak disorder, which matches with the prediction of the Ginibre random matrices. At the high disorder strength, the disappearance of the ramp indicates uncorrelated energy levels and thereby localization in the non-hermitian Hamiltonian. The inset shows the DSFF at $\phi=0$ for disorder strength $h=2$ which exactly matches with $2K_c^{\mathrm{GinUE}}$.}
\label{fig:dsff-1}
\end{figure}

Fig.~\ref{fig:dsff-1} shows the DSFF (Eq.~\ref{eq:dsff}) as a function of the rescaled time variable $|\tilde{\tau}|$ for model-I for small ($h=2$) and large disorder strength ($h=14$). The rescaled time variable is defined as $\tilde{\tau}=\tau/\tau_H$ where $\tau_H$ is the Heisenberg time chosen so that the time axis is shifted appropriately to facilitate an agreement with the RMT behavior (see appendix-\ref{appendix-c} for details). For small disorder strength ($h=2$), at  very small time scales $|\tilde{\tau}|$, the DSFF starts from a value $\mathcal{N}$. After initial oscillatory behavior, the DSFF for model-I has a non-linear ramp characteristic of the Ginibre random matrices indicating chaotic behavior. At very large time scales, the DSFF saturates to $1$. For some intermediate time $\tilde{\tau}$ before the onset of ramp, DSFF $K(\tilde{\tau},\tilde{\tau}^*)$ depends on both $|\tilde{\tau}|$ and $\phi$, and this is an artifact of the absence of rotational symmetry in the distribution of $z_{mn}$ (see appendix-\ref{appendix-a}). The universal non-linear ramp obtained in model-I agrees with the analytical expression of DSFF for  GinUE, given by~\cite{dsff_Li.2021}
\begin{equation}\label{eq:ginue}
    K^{\mathrm{GinUE}}_{c}(\tilde{\tau},\tilde{\tau}^*)=1-\mathrm{exp}\left(-|\tilde{\tau}|^2/4\right).
\end{equation}
Here, the rescaled time variable is defined by $\tilde{\tau}=\tau/\tau_H$ where $\tau_H=\sqrt{\mathcal{N}}$ is the Heisenberg time for GinUE random matrices and is inversely proportional to the mean level spacing ($\sim 1/\sqrt{\mathcal{N}}$).

Since the Hamiltonian of model-I is completely real and thus has complex conjugation symmetry ($H=H^*$), its behavior in the quantum chaotic regime should correspond to Ginibre orthogonal ensembles (GinOE)~\cite{Ginibre.1965,kawabata2019,universality_hamazaki.2020} instead of GinUE. However, just like level spacing distribution, DSFF also has the same form for all three Ginibre symmetry classes except for some special angles near $\phi=0,\pi/2$~\cite{dsff_Li.2021} for which the analytical expression is not known. A non-hermitian matrix belonging to GinOE has  a eigenvalue spectrum comprising of complex conjugate pairs and a significant number of completely real eigenvalues. These give rise to accidental degeneracies when projected onto $\phi=0,\pi/2$ directions. Because of these degeneracies at $\phi=0$, the DSFF for GinOE is exactly twice the DSFF for GinUE ($K_c^{\mathrm{GinOE}}|_{\phi=0}=2K_c^{\mathrm{GinUE}}$) in the limit of large matrix size. Thus, for generic values of $\phi$ away from $0$ and $\pi/2$, it is justified to compare the DSFF of GinOE class with that of GinUE. On the other hand, for large disorder strength ($h=14$), the DSFF saturates to $K=1$ after initial dip and the absence of a ramp implies uncorrelated energy levels. We find that for intermediate values of $h$, especially in the proximity of the critical point for the localization transition~\cite{NHMBL_hamazaki.2019}, the data neither resembles RMT nor Poisson statistics. This is rooted in finite size effects and details are discussed in appendix-\ref{appendix-d}.
\begin{figure}[t]
\centering
\includegraphics[width=0.5\textwidth]{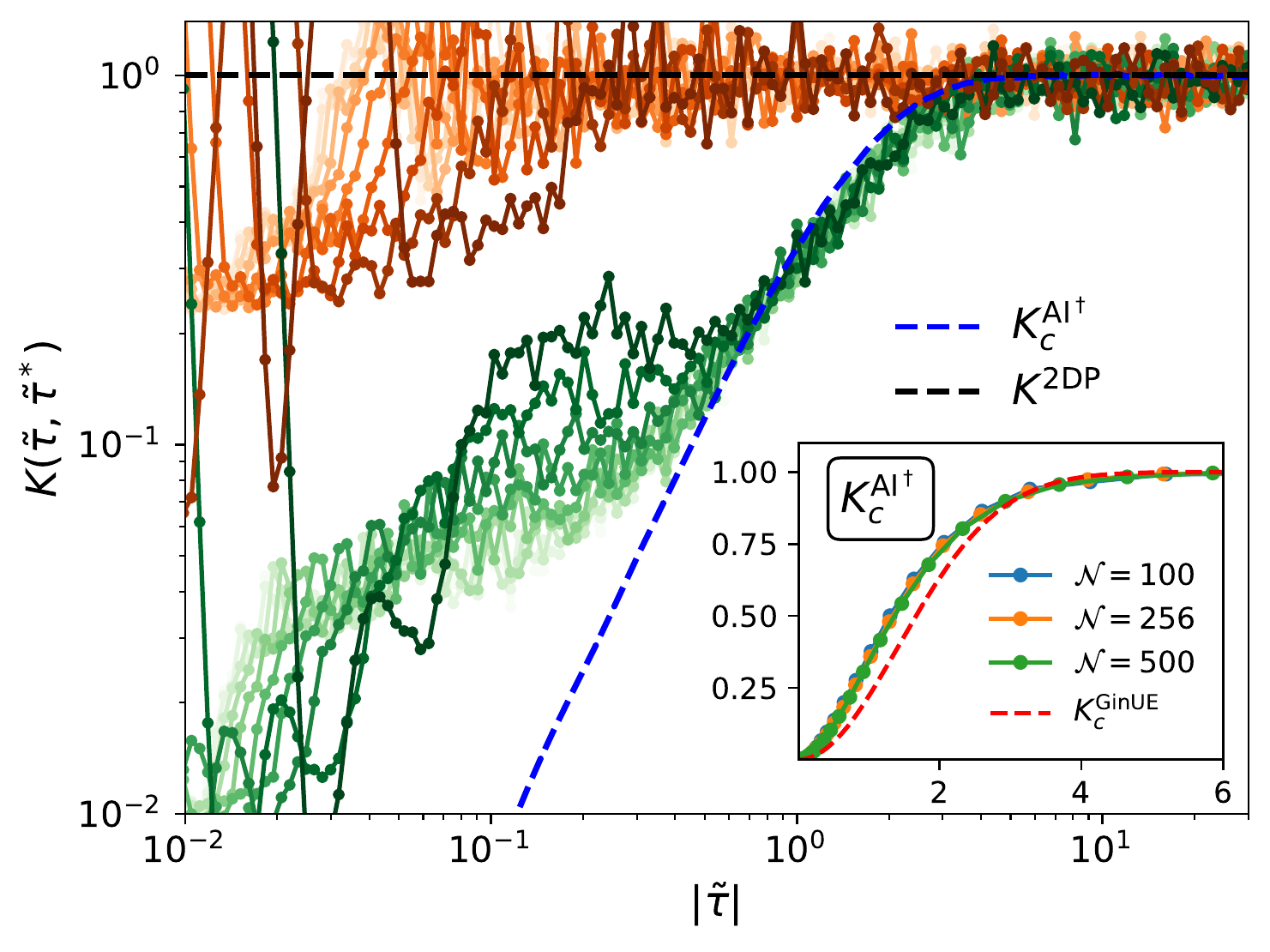}
\caption{[Model-II, Eq.~\ref{eq:model2}] DSFF for model-$II$: Dissipative spectral form factor for model $II$ as a function of rescaled time variable $|\tilde{\tau}|$ for different values of $\phi \in [0, 9\pi/20]$ in steps of $\pi/20$ for $h=2$ (Green), $14$ (Orange). The lowest value of $\phi$ in the given range corresponds to the lightest shade, while the highest value of $\phi$ corresponds to the darkest shade of the corresponding colors. The blue line shows DSFF for random matrices belonging to the $\mathrm{AI}^\dagger$ symmetry class, while the black line corresponds to the uncorrelated random complex energy levels. The inset shows the DSFF for random matrices of size $\mathcal{N}\times \mathcal{N}$ belonging to the $\mathrm{AI}^\dagger$ symmetry class where the DSFF for GinUE is plotted as a red dashed line to highlight the difference between $\mathrm{AI}^\dagger$ and GinUE symmetry classes.}
\label{fig:dsff-2}
\end{figure}

Fig.~\ref{fig:dsff-2} shows the variation of DSFF as a function of the rescaled time variable $|\tilde{\tau}|$ for model-II [Eq.~\ref{eq:model2}]. This has the same qualitative feature as in model-I. The ramp here does not match with the GinUE case as the symmetry of the Hamiltonian does not conform with any of the Ginibre classes. The Hamiltonian of this model has a transposition symmetry ($H=H^T$) similar to $\mathrm{AI}^\dagger$ symmetry class of non-hermitian matrices~\cite{kawabata2019, universality_hamazaki.2020}. We, therefore, calculate the DSFF for non-hermitian random matrices belonging to the $\mathrm{AI}^\dagger$ symmetry class (shown in the inset of Fig.~\ref{fig:dsff-2}) and compare it with the DSFF obtained for model-II at weak disorder. We find good agreement of the DSFF of model-II with that of $\mathrm{AI}^\dagger$ symmetry class at weak disorder, while at large disorder strength, it matches with that of uncorrelated complex energy levels.
\begin{figure}[t]
    \centering
    \includegraphics[width=0.5\textwidth]{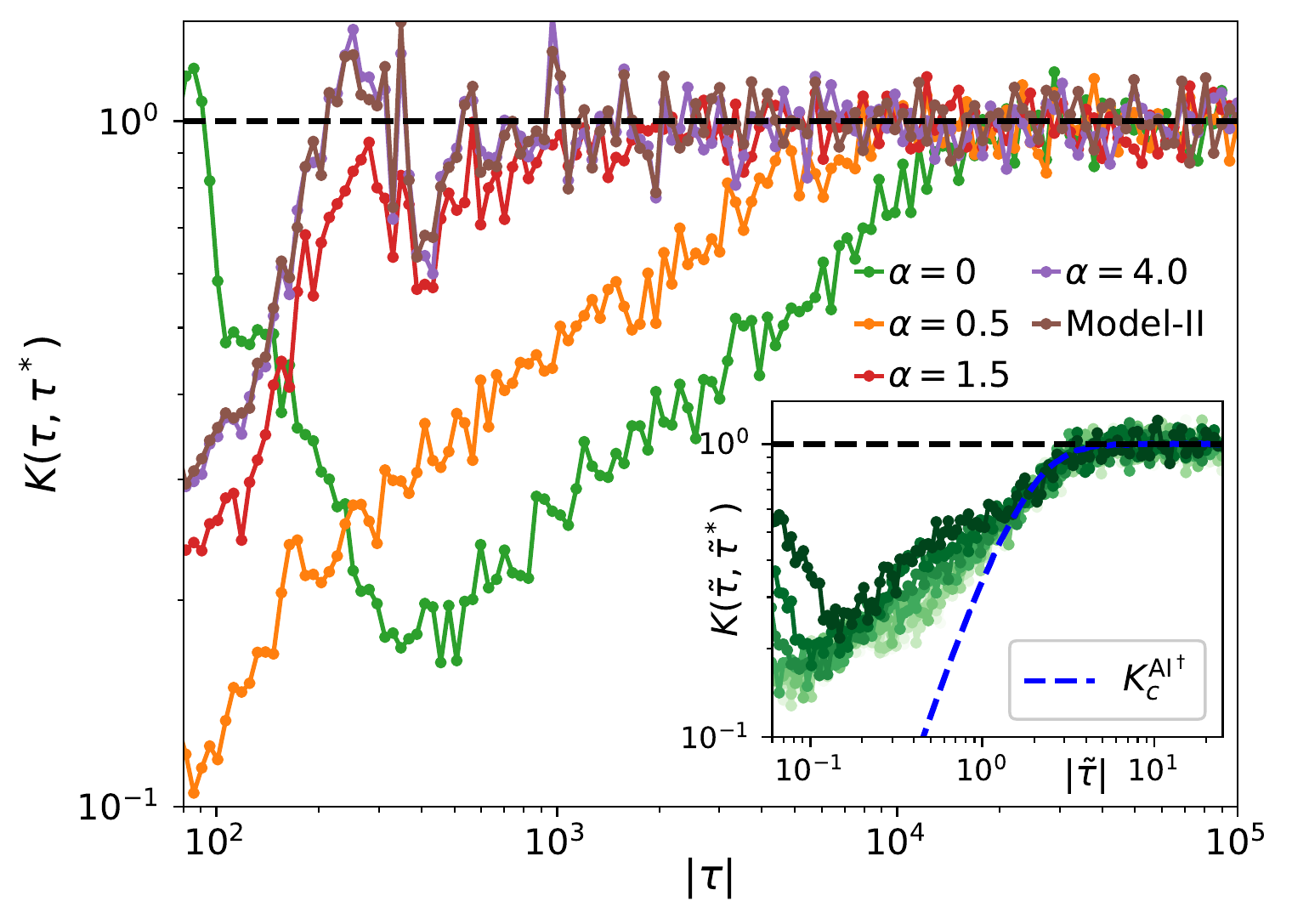}
    \caption{[Model-III, Eq.~\ref{eq:model3}] DSFF for model-III as a function of the time variable $|\tau|$ for different values of $\alpha$  at $\phi=0.3\,\pi$ and system size $L=16$. The disorder strength is fixed at $h=14$. The DSFF for model-II having only nearest-neighbor hopping $(\alpha\rightarrow\infty)$ is also plotted in the same figure for comparison. The DSFF changes from that of uncorrelated complex levels (black dashed line) to that of $\mathrm{AI}^\dagger$ symmetry class (shown in inset) as the value of $\alpha$ is decreased. The inset shows agreement of $K_c^{\mathrm{AI}^\dagger}$ with the $\alpha=0$ case, where different shades of the green color represents different values of $\phi$ chosen from $[0,9\pi/20]$ in steps of $\pi/20$.}
    \label{fig:dsff-3}
\end{figure}

Next, we discuss the effect of introducing long-range hopping in the model-II (called model-III [Eq.~\ref{eq:model3}]). In Fig.~\ref{fig:dsff-3} we show the DSFF for model-III. We tune the parameter $\alpha$ that controls the range of the hopping. While doing that, we fix the disorder strength at $h=14$ such that in the presence of only short range hopping (nearest-neighbor) the DSFF is similar to that obtained from uncorrelated random complex levels. From Eq.~\ref{eq:model3}, it is evident that $\alpha\rightarrow \infty$ corresponds to a system comprising of only nearest-neighbor hopping. We see that as $\alpha$ is decreased, thereby increasing the range of hopping for the particles, the system becomes more chaotic. There is good agreement of the DSFF with that of $\mathrm{AI}^\dagger$ symmetry class when $\alpha=0$ as shown in the inset of Fig.~\ref{fig:dsff-3}. This is owing to the fact that changing the range of the hopping does not change the symmetry of the Hamiltonian, therefore, implying that the Hamiltonian of model-III preserves the transposition symmetry ($H=H^T$). For intermediate values of $\alpha$, the DSFF results neither fall into the RMT regime nor correspond to the Poisson statistics. This is again rooted in the finite size effects. The enhancement of the chaotic signatures with the increasing range of hopping (decreasing $\alpha$) in this non-hermitian Hamiltonian complements the understanding in the case of long-range Hermitian Hamiltonians, where the delocalization increases with increase in the range of hopping~\cite{Nag_longrange.2019,Roy_longrange.2019}.

All the DSFF results presented above for the three models were for the case when no unfolding procedure was involved. We find that one of the proposed unfolding procedures for such a complex eigen-spectrum, namely the conformal mapping~\cite{Conformal.1999}, yields the same DSFF. In addition to DSFF, which has been proven to be a promising diagnostic for identifying the long range correlations of the complex eigenvalues, we discuss another recently introduced quantity, namely the complex spacing ratio, which is the complex generalization of the level spacing ratio used widely in the numerical calculations of Hermitian Hamiltonians~\cite{Oganesyan.2007}. Needless to mention this quantity is insensitive to any unfolding procedure or non-universal features of underlying density of states.
\subsection{Complex spacing ratio (CSR)}
The complex spacing ratio for the $n^{\mathrm{th}}$ eigenvalue is defined as the ratio of complex differences given by
\begin{equation}\label{eq:csr}
\xi_{n} = \frac{z_{n}^{NN}-z_{n}}{z_{n}^{NNN}-z_{n}} = r_{n} e^{i \theta_n}
\end{equation}
where $z_n^{NN}$ and $z_n^{NNN}$ are the nearest and the next nearest neighbor of the energy level $z_n$ respectively~\cite{cspacing_lucas.2020}. Here the notion of distance is the absolute distance between eigenvalues in the complex plane. Note that $r_n$ and $\theta_n$ are respectively the absolute value and the argument of  the complex ratio $\xi_n$ defined in Eq.~\ref{eq:csr}.
The nearest-neighbor difference depends on the position of the energy level in the complex spectrum, hence on the local density of states. However, in the ratio $\xi_n$, the effect of the local density of states is washed away thereby making CSR a highly preferable diagnostic which is insensitive to the unfolding procedures. Note from the definition of CSR that $r_n \in [0,1]$ and $\theta_n \in [-\pi,\pi] \quad \forall \quad n$. Here, the quantities of interest are the marginal distributions of $r$, denoted by $\rho(r)$, and $\theta$, denoted by $\rho(\theta)$,  where the distributions are obtained considering all energy levels.

Both the distributions $\rho(r)$ and $\rho(\theta)$ have different behavior for different random matrix ensembles (see appendix-\ref{appendix-b})~\cite{cspacing_lucas.2020}.
In the presence of chaos, the energy levels experience level repulsion, and $\rho(r)$ vanishes for small $r$. It is worth noting that similar to level spacing distributions and DSFF, all three Ginibre ensembles (GinOE, GinUE, GinSE) have the same form for the marginal distributions of $r$ and $\theta$.
On the other hand, for complex uncorrelated energy levels, the CSR is uniform inside a unit circle with the marginal distributions given by $\rho(r)=2 r$ and $\rho(\theta)=1 /(2 \pi)$~\cite{cspacing_lucas.2020}.
Thus the marginal distributions can serve as excellent diagnostics both in the chaotic and localized phase. We calculate both the marginal distributions for different disorder strengths considering all the energy levels in the complex spectrum for all disorder samples. In addition to the marginal distributions, the averages $\langle r \rangle$ and $-\langle \mathrm{cos} \theta\rangle$ have different values for different random matrix ensembles (tabulated in appendix-\ref{appendix-e}). These values are very different from the case of uncorrelated complex energy levels where $\langle r\rangle$ = 2/3 and $\langle\cos \theta\rangle=0$. The analytical expressions for the marginal distributions for the Ginibre random matrices are not known. Recently, attempts have been made to derive approximate analytical expressions for these distributions~\cite{approx_ginibre}.

\begin{figure}[t]
\centering
{\includegraphics[width=0.48\textwidth]{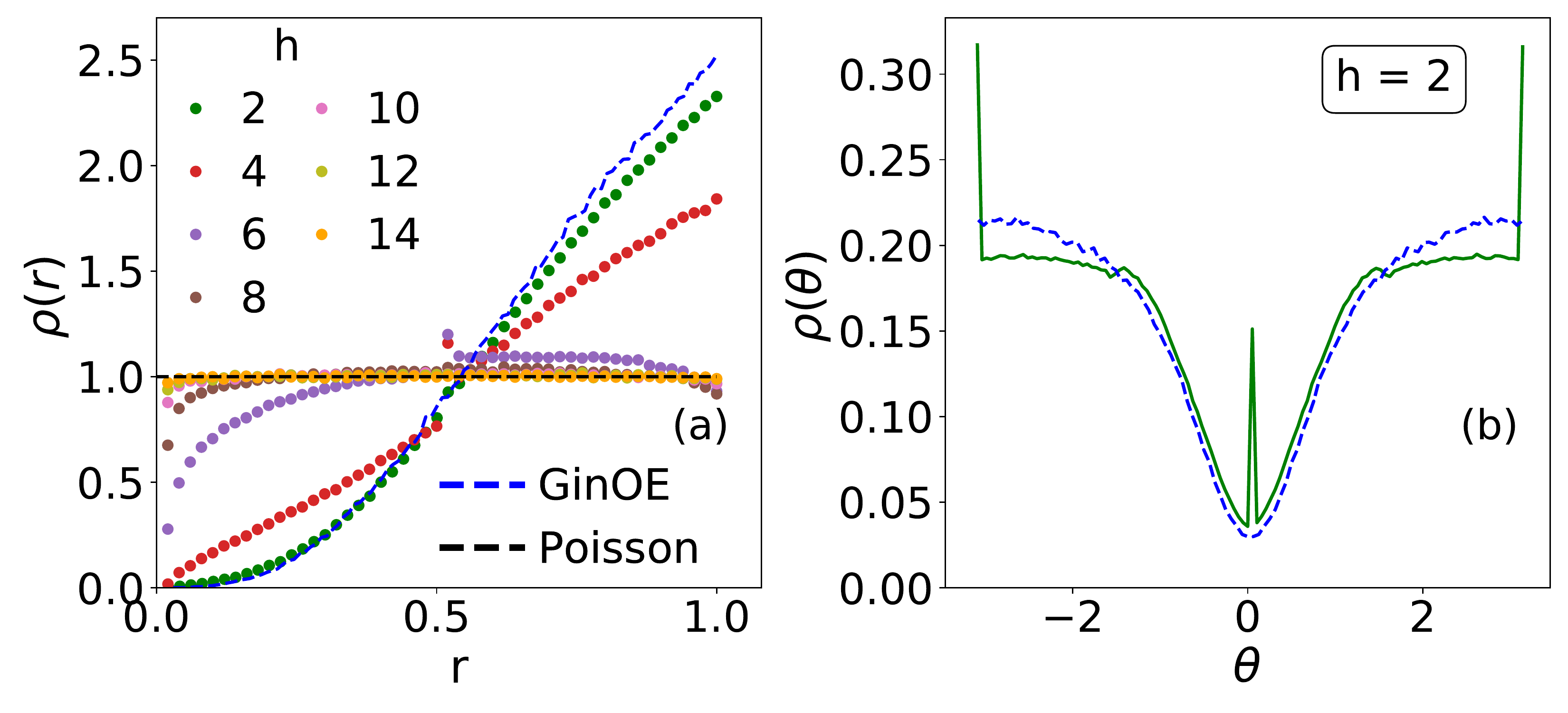}}
\caption{[Model-I,Eq.~\ref{eq:model1}] (a) Marginal distribution of $r$ for different values of $h$ for $L=18$. For weak disorder strength ($h=2$) the distribution corresponds to that of GinOE (blue dashed line), while for large $h$ we see agreement with 1D Poisson statistics (black dashed). (b) The marginal distribution of $\theta$ for $h=2$ shows agreement with that of GinOE (blue dashed line) while the peaks at $0,\pi$ are due to the completely real eigenvalues. The variable $\theta$ becomes ill-defined at large disorder strength due to the completely real spectrum.}
\label{fig:112}
\end{figure}

Fig.~\ref{fig:112} (a) represents the marginal distributions of $r$ for model-I, as the disorder strength $h$ is varied.
Model-I has a complex-real transition as a function of disorder strength $h$ where after some critical value of $h$ the spectrum becomes completely real~\cite{NHMBL_hamazaki.2019}.
Despite this complex-real transition, we use the same definition of CSR (Eq.~\ref{eq:csr}) even for real eigenvalues in the strong disorder limit in order to facilitate the use of the same diagnostic for all disorder strengths. In the presence of strong disorder, where we expect uncorrelated energy levels from DSFF results (Fig.~\ref{fig:dsff-1}), the CSR ($\xi_n$) is real due to the real energy spectrum. In fact, for real uncorrelated energy levels, the CSR turns out to be uniformly distributed between $-1$ and $1$. Hence, the absolute values of the ratio given by $r$ are uniformly distributed between $0$ and $1$ and $\langle r\rangle=1/2$. In other words, $\rho(r)=\Theta(1-r)$~\cite{cspacing_lucas.2020}.
 
We find that in the presence of weak disorder($h=2$), the marginal distribution of $r$ corresponds to that of GinOE. As the disorder strength $h$ is increased, the marginal distribution of $r$ starts becoming more and more flat and eventually corresponds to 1D Poisson statistics, i.e., $\rho(r)=1$ demonstrating the localization transition. Since we consider the entire spectrum for our results, we see some deviation from GinOE for larger values of $r$ for $h=2$ as shown in Fig.~\ref{fig:112}. Upon suitably going to the middle of the spectrum, as is usually done~\cite{Oganesyan.2007,NHMBL_hamazaki.2019}, the agreement between the distribution of $r$ and that of GinOE becomes better as shown in appendix-\ref{appendix-e} (Fig.~\ref{fig:11}). Note that in Fig.~\ref{fig:11}, there is still a discrepancy due to a small fraction of completely real eigenvalues of model-I, which are comparatively much less in GinOE random matrices. Since model-I undergoes complex-real transition as the disorder strength is increased, the marginal distribution of the angle $\theta$ is not particularly helpful. This is because $\theta$ can have only two values $0$ and $\pi$. Nonetheless, for weak disorder strength ($h = 2$), the distribution of $\theta$ is well defined and agrees with that of GinOE [Fig.~\ref{fig:112}(b)].

\begin{figure}[t]
\centering
\includegraphics[width=0.48\textwidth]{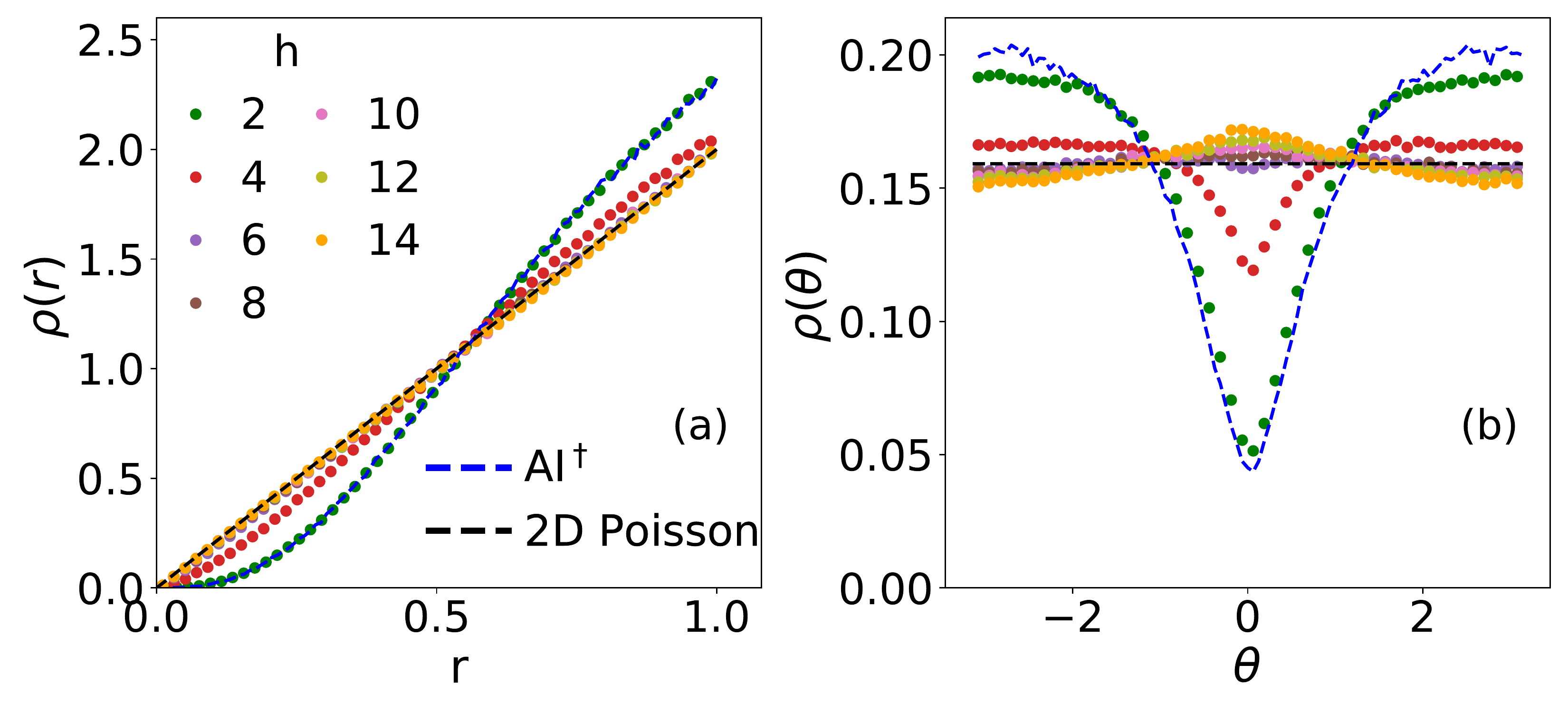}
\caption{[Model-II, Eq.~\ref{eq:model2}] Marginal distributions of (a) $r$ and (b) $\theta$ for different values of $h$ for $L=18$. For smaller disorder strength ($h=2$), both the marginal distributions correspond to that of AI$^\dagger$ symmetry class (blue dashed), while for large disorder strength $h$ they correspond to that of 2D Poisson statistics (black dashed).}
\label{fig:121}
\end{figure}
\begin{figure}[h!]
\centering
\includegraphics[width=0.48\textwidth]{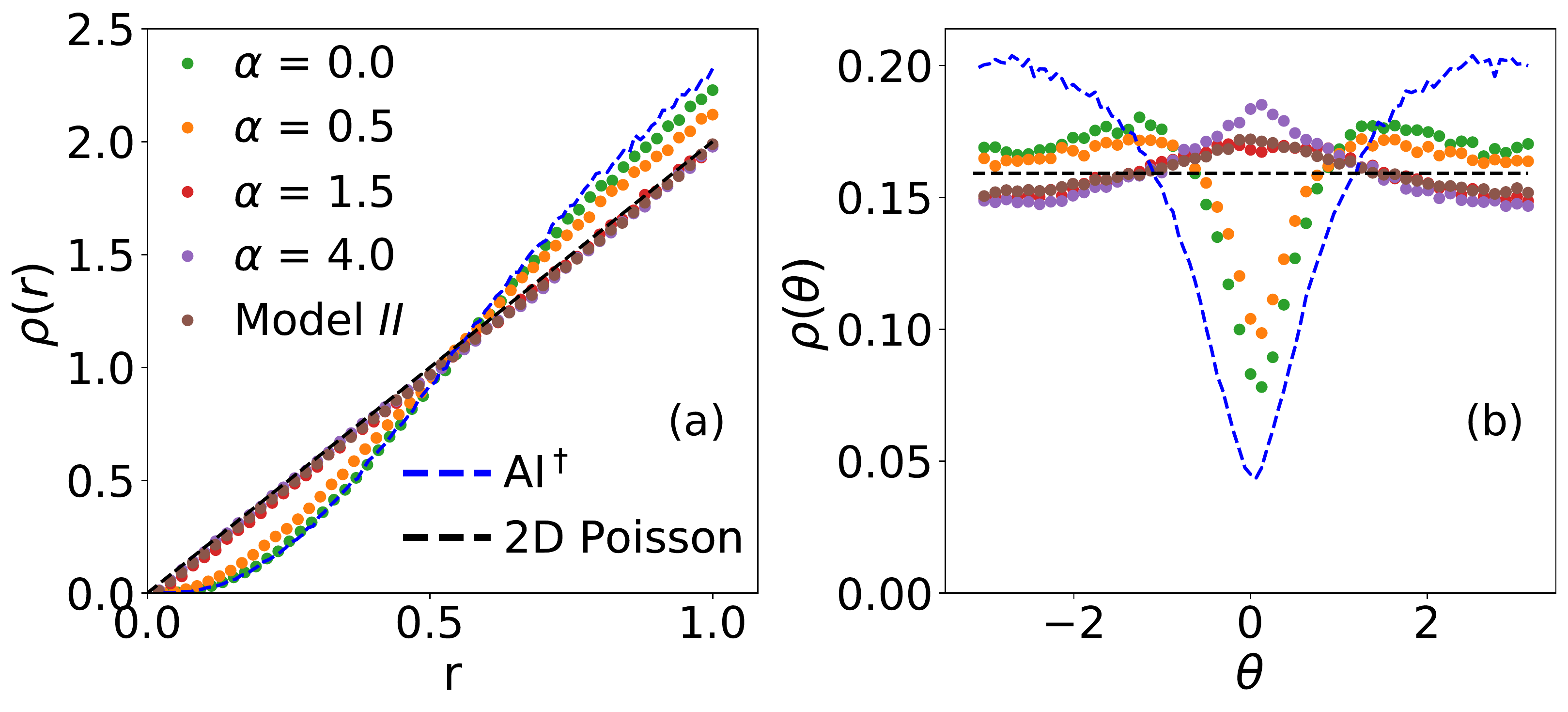}
\caption{[Model III, Eq.~\ref{eq:model3}] Marginal distributions of (a) $r$ and (b) $\theta$ for different values of $\alpha$ at $h=14$ for system size $L=16$. Both the marginal distributions change from that of 2D Poisson statistics (black dashed) to that of  $\mathrm{AI}^\dagger$ symmetry class (blue dashed)  as $\alpha$ decreases. In addition, we also plot the marginal distribution for model-II at $h=14$ (which corresponds to $\alpha\rightarrow\infty$).}
\label{fig:124}
\end{figure}
Fig.~\ref{fig:121} shows the variation of marginal distribution of $r$ and $\theta$ as the disorder strength $h$ is increased for model-II. For small disorder strength ($h = 2$) both the distributions correspond to that of AI$^{\dagger}$ symmetry class. As the disorder strength $h$ is increased, both the marginal distributions tend to those corresponding to 2D Poisson statistics. In this case too, if one restricts the analysis to the middle of the spectrum the agreement with random matrix theory becomes better as shown in appendix-\ref{appendix-e} (Fig.~\ref{fig:19}). Fig.~\ref{fig:124} shows the marginal distribution of $r$ and $\theta$ for model-III at disorder strength $h=14$ as $\alpha$ is varied. Here too, the agreement with random matrix theory becomes better taking just the middle of the spectrum (appendix-\ref{appendix-e}, Fig.~\ref{fig:app-rho-3}). Similar to DSFF, we notice that the behaviors of both the distributions $\rho(r)$ and $\rho(\theta)$ change from the case of 2D Poisson statistics to that of $\mathrm{AI}^\dagger$ symmetry class as the range of hopping is increased by decreasing $\alpha$.

For intermediate values of $h$ (or $\alpha$ for model-III), in all the three models discussed above, both the DSFF and the CSR show neither perfect agreement with RMT nor Poisson statistics. This is owing to the finite size effects in the vicinity of the critical point for the localization transition as elaborated in appendix-\ref{appendix-d}. The marginal distributions $\rho(r)$ and $\rho(\theta)$ tend towards the expected RMT/ Poisson statistics as one increases the system size. It is worth mentioning that in addition to the marginal distributions, the average values of $r$ and $\mathrm{cos}~\theta$ can be computed and are summarized in tables (Table~\ref{101},\ref{102},\ref{103}) and figures(Fig.~\ref{fig:raverage-I},\ref{fig:raverage-II}) (see appendix-\ref{appendix-e}).
\section{Conclusions and Outlook}\label{sec:conclusion}
In this work, we discuss spectral properties of three different non-hermitian Hamiltonians (Eqns.~\ref{eq:model1}, \ref{eq:model2}, \ref{eq:model3}) each of which are unique in their own way. We discuss two quantities namely dissipative spectral form factor (DSFF) and complex spacing ratio (CSR) which serve as an excellent diagnostic to classify phases. For the two models with short-ranged hopping, we show that both the quantities DSFF (Figs.~\ref{fig:dsff-1},~\ref{fig:dsff-2}) and CSR (Figs.~\ref{fig:112},~\ref{fig:121}), capture the chaotic (localizing) behavior at the weak (strong) disorder limit. In the chaotic regime, both the quantities show agreement with that of the respective non-hermitian random matrix ensembles despite the inhomogeneous and anisotropic distribution of the energy levels in the complex plane. In other words, both the quantities show universal features in the chaotic regime despite local non-universal properties of the density of states. On the other hand, in the presence of strong disorder, we find that all the energy levels are uncorrelated and follow Poisson statistics.

We also discuss long-ranged generalization for one of the models. We show that with the increasing range of hopping, the system becomes more and more chaotic (Figs.~\ref{fig:dsff-3},~\ref{fig:124}). A disorder strength which is strong enough for generating uncorrelated energy levels in the case of short-range models becomes insufficient and energy level correlations arise in the presence of long-range connectivity. In fact in the limit of all-to-all coupling with the same hopping parameter strength, the DSFF and CSR correspond to that of RMT in spite of the disorder strength being quite large as compared to the other scales of the problem.

In the chaotic regime, the DSFF has a dip-ramp-plateau structure with a non-linear ramp characteristic to non-hermitian random matrices~\cite{dsff_Li.2021}. At intermediate time scales ($\tau\lesssim 1$) before the universal ramp, where the unconnected part is already zero, the DSFF depends on $\phi$ as well as system size. This naturally raises a discussion of Thouless time, which is defined as the time scale after which the quantum dynamics of the system is governed by the random matrix theory. The Thouless time $t_{th}$~\cite{Edwards.1972}, which was initially used in the scaling theory of Anderson Localization~\cite{Abrahams.1979,Lee.1985,Kramer.1993}, has been used in interacting systems to understand the nature of the many-body localization transition~\cite{Abanin-tth.2021,Suntajs-tth.2020,Sierant-tth.2020,Sonner-tth.2021}.
In our calculation, a time scale can be defined before which the DSFF depends on the system size as well as the angle $\phi$. However, the relation of this time scale with a Thouless time like quantity for such non-hermitian systems~\cite{shivam2022many} is far from being fully understood and remains an interesting direction for future investigation.

The possible link between spectral properties of non-hermitian disordered systems to other quantities such as commonly computed imbalance remains an interesting question. Although in this work we considered conventional random disordered potential, such a thorough investigation for non-hermitian quasiperiodic systems could be important.
The crossover from localization to chaos as one changes the range of hopping in interacting non-hermitian disordered networks is a new direction and warrants further detailed investigation.
\section*{Acknowledgements}
We thank Mahaveer Prasad, Hari Kumar Yadalam for helpful discussions.
M.K. acknowledges the support of the Ramanujan Fellowship (SB/S2/RJN-114/2016), SERB Early Career Research Award (ECR/2018/002085) and SERB Matrics Grant (MTR/2019/001101) from the Science and Engineering Research Board (SERB), Department of Science and Technology, Government of India. M.K. acknowledges support of the Department of Atomic Energy, Government of India, under Project No. RTI4001.  We gratefully acknowledge the ICTS-TIFR high performance computing facility.  M.K. thanks the hospitality of  \'Ecole Normale Sup\'erieure (Paris). M.K. acknowledges support from the Infosys Foundation International Exchange Program at ICTS. 
\appendix
\section{Distribution of complex energy differences}\label{appendix-a}
\begin{figure}[h!]
\centering
\includegraphics[width=0.48\textwidth]{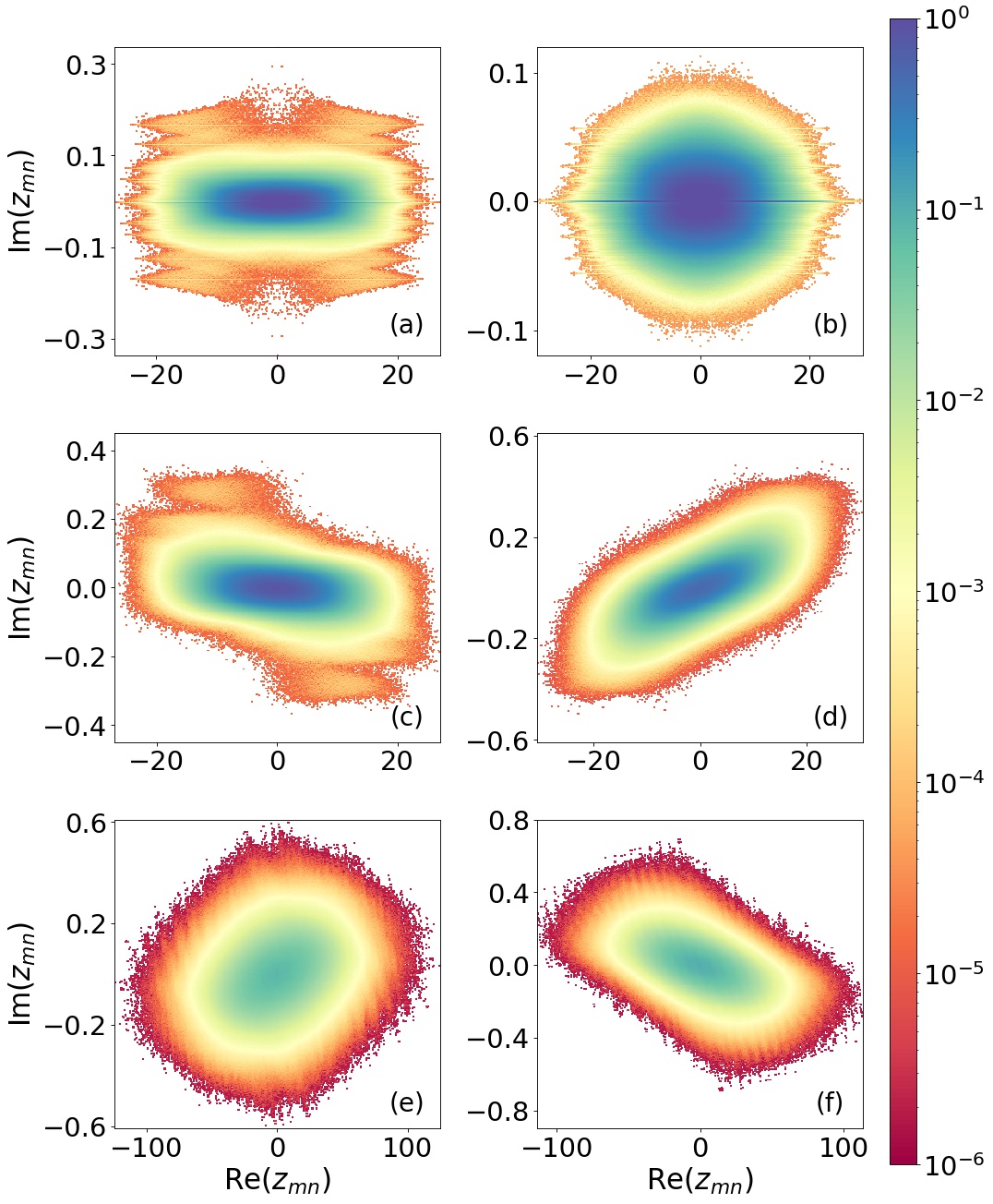}
\caption{Distribution of complex energy difference considering two separate disorder samples each for Model-I [(a),(b)] ($h = 2$), Model-II [(c),(d)] ($h = 2$) and Model-III ($h = 14, \alpha=0$) [(e),(f)] at system size $L=16$. The color map shows the density of the energy differences $z_{mn}$ in the complex plane. This clearly shows the lack of rotational symmetry (anisotropy) unlike non-hermitian random matrix ensembles.}
\label{fig:zmn_distr}
\end{figure}

In this appendix, we discuss the distribution of the difference in complex energies $z_{mn}=z_m-z_n$. This analysis is of pivotal importance since it is the main ingredient in dissipative spectral form factor (DSFF) which captures all long-range correlations of complex eigenvalues [Eq.~\ref{eq:dsff}]. In Fig.~\ref{fig:zmn_distr}, we show the distribution of $z_{mn}$ for two distinct representative samples in the complex plane for all the three models in the chaotic regime. The color map shows the density of these energy differences. Some features of this color map are worth elaborating further.

The energy spectra for the disordered systems in the complex plane is inhomogeneous and anisotropic. This is also reflected in the complex energy differences as seen in Fig.~\ref{fig:zmn_distr}. As a consequence the DSFF as a function of $|\tau|$ depends on the choice of the angle $\phi=\mathrm{arg}(\tau)$. This is reflected in the early time behavior of the DSFF after the initial dip. At intermediate times, the DSFF is robust to this $\phi$ variation thereby leading to a universal ramp feature. The late time behavior is manifested as a plateau which is $\phi$ independent as expected. Such $\phi$ dependency in DSFF which stems from the lack of rotational symmetry in the energy spectrum and energy differences is absent in non-hermitian random matrix ensembles and certain non-hermitian models~\cite{dsff_Li.2021}.
\section{Non-hermitian random matrices belonging to $\mathrm{AI}^\dagger$ symmetry class}\label{appendix-b}
\begin{figure}[t]
\centering
\includegraphics[width=0.48\textwidth]{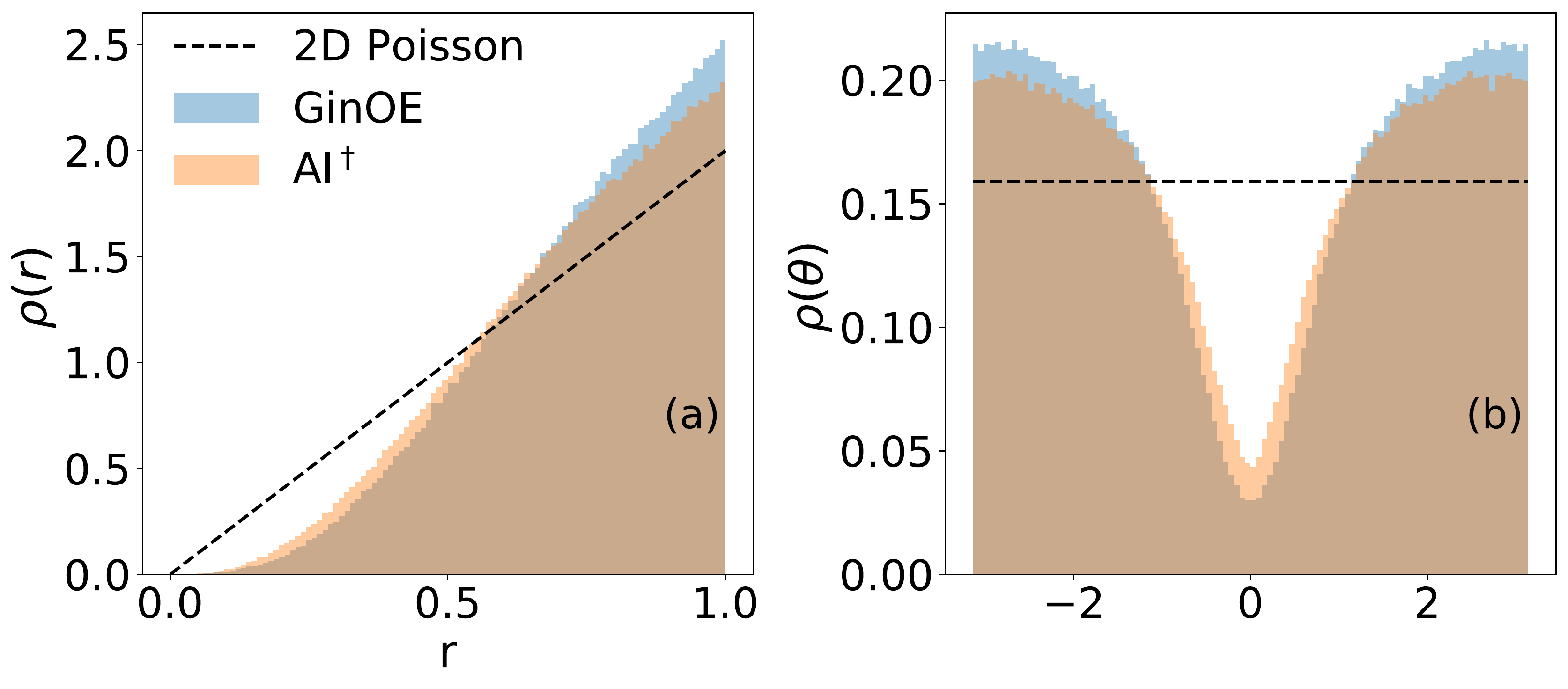}
\caption{Marginal distributions of (a) $r$ and (b) $\theta$ for GinOE (blue), AI$^\dagger$ (orange) symmetry class and 2D Poisson statistics (black dashed).}
\label{fig:aid}
\end{figure}

Here, we describe the calculation of DSFF and CSR for the non-hermitian random matrices belonging to $\mathrm{AI}^\dagger$ symmetry class. The non-hermitian random matrices belonging to the $\mathrm{AI}^\dagger$ symmetry class~\cite{kawabata2019, universality_hamazaki.2020}. has the transposition symmetry ($H=H^T$). A random matrix with such symmetry has random complex numbers as matrix elements with the constraint $H_{ij}=H_{ji}$. Both the real and imaginary parts of the elements $H_{ij}$ $(i\geq j)$ are independent and identically chosen from a Gaussian distribution $\mathbb{N}(0,1/\sqrt{2})$. The complex energy spectrum of such a matrix of dimension $\mathcal{N}\times \mathcal{N}$ is a uniform disc of radius $R=\sqrt{\mathcal{N}}$~\cite{forrester2010log}. Thus, the energy spectrum of a matrix $H'=H/\sqrt{\mathcal{N}}$ is a disc of unit radius. We implement Eq.~\ref{eq:dsff} and Eq.~\ref{eq:csr} to calculate the DSFF and CSR respectively and compare both the quantities with known results of GinOE random matrices. Both the quantities show a different functional form from that of GinOE random matrices as depicted in Fig.~\ref{fig:dsff-2} (inset) and Fig.~\ref{fig:aid}.
\section{Estimation of Heisenberg time}\label{appendix-c}
Here we discuss the calculation of Heisenberg time for the models. For the DSFF of model-I to agree with the analytical form of the DSFF for GinOE random matrices given by~Eq.~\ref{eq:ginue}, we define the Heisenberg time scale $\tau_H$ which scales the bare time $\tau$ as $\tilde{\tau}=\tau/\tau_H$. To estimate $\tau_H$, we first locate the time scale  $\tau^*$ after which the DSFF shows universal behavior (non-linear ramp) and fit the curve \begin{equation}\label{eq:fitting}
f_K(\tau,m)=1-\mathrm{exp}\left[-m|\tau|^2\right]
\end{equation}
with the  numerical data of the DSFF where $m$ is the fitting parameter and $\tau_H$ is extracted as $\tau_H=\frac{1}{2\sqrt{m}}$. For better accuracy, we calculate $\tau_H$ for different choices of $\phi$ and take the average over $\phi$ to calculate the Heisenberg time used for rescaling the time axis in Fig.~\ref{fig:dsff-1}. In the case of model-II and model-III, we compare the numerical data with the DSFF of $\mathrm{AI}^\dagger$ symmetry class for which the analytical expression is not known yet. However, we use the same fitting function~Eq.~\ref{eq:fitting} to get an estimate of $\tau_H$ (just like model-I) and use that to shift the numerical result for $\mathrm{AI}^\dagger$ symmetry class to show agreement between model-II/model-III with that of $\mathrm{AI}^\dagger$ symmetry class. This procedure of estimating $\tau_H$ works well
because the DSFF for GinOE and $\mathrm{AI}^\dagger$ symmetry class have common basic features despite having different functional variations as depicted in the inset of Fig.~\ref{fig:dsff-2}.
\begin{figure}[h!]
    \centering
    \includegraphics[width=0.48\textwidth]{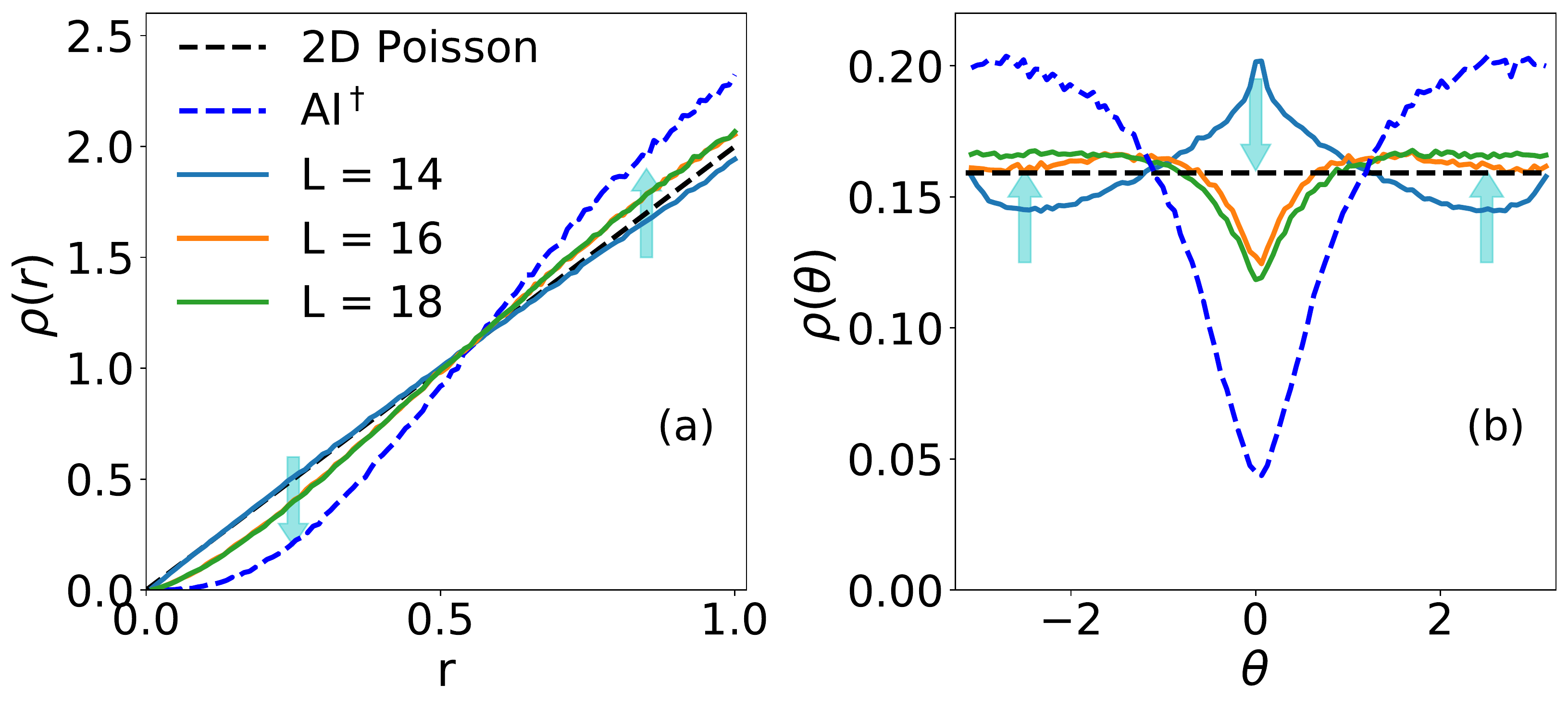}
\caption{[Model-II, Eq.~\ref{eq:model2}] Marginal distributions for (a) $r$ and (b) $\theta$ for $h = 4$ for different system sizes $L$ for entire spectrum statistics. We find evidence of deviations from the expected RMT behavior due to finite size effect. In other words, upon increasing system size, we notice that our results approach that of marginal distributions for $\mathrm{AI}^\dagger$ symmetry class.}
    \label{fig:h4_finitesize}
\end{figure}
\section{Finite-size effects in complex spacing ratio}\label{appendix-d}
In this appendix, we discuss the deviation of the complex spacing ratio (CSR) from the RMT predictions for the models at the disorder strengths corresponding to the intermediate region between the chaotic and the localized regimes. In the main text, the DSFF and the CSR for different models are shown deep in the chaotic and the localized regime, where there are good agreement with the corresponding RMT/ Poisson statistics at the system size $L=18$. However, this does not hold for intermediate disorder strengths near the critical point for the localization transition, where both the DSFF and CSR are prone to finite size effects. 

Fig.~\ref{fig:h4_finitesize} shows the variation of both the marginal distributions of $r$ and $\theta$ for different system sizes at $h=4$ for model-II, which corresponds to the intermediate disorder strength for this model. For system size, $L=14$, both the marginal distributions show similarity with the 2D Poisson statistics implying uncorrelated random complex energy levels. This, in fact, turns out to be a finite size effect, and as the system size increases both the quantities deviate from the Poisson statistics and tend to approach the statistics of the $\mathrm{AI}^\dagger$ symmetry class. However, it never reaches the statistics of this symmetry class for the choice of system sizes making the analysis inconclusive for intermediate disorder strengths. Similarly, for other intermediate values of disorder strength $h$ for all the models, finite size effects can be visible both in marginal distribution and DSFF, which makes the identification of the critical parameter for the chaotic-localization transition difficult.
\begin{widetext}
\begin{center}
\begin{table}[h!]
	\begin{tabular}{| l | l | l | l | l | l | l | l | l | l |}
	\hline
	&\multirow{2}{*}{GinOE}&\multicolumn{7}{|c|}{h} & \multirow{2}{*}{1D P} \\ \cline{3-9}
	&&$2$&$4$&$6$&$8$&$10$&$12$&$14$& \\ \hline
	$-\langle \mathrm{cos} \theta \rangle$ & $0.244$ & $0.196$ & $0.018$& $0.023$ & {\Large $\times$} & {\Large $\times$} & {\Large $\times$} & {\Large $\times$} & {\Large $\times$} \\ \hline
	$\langle r \rangle$ & $0.738$ & $0.736$ & $0.703$& $0.553$& $0.508$& $0.502$& $0.501$& $0.5$ & $1/2$ \\ \hline
	$-\langle \mathrm{cos} \theta \rangle_M$ & $0.244$ &$0.228$& $0.039$& $0.019$ & {\Large $\times$} & {\Large $\times$} & {\Large $\times$} & {\Large $\times$} &  {\Large $\times$}\\ \hline
	$\langle r \rangle_{M}$ & $0.738$&$0.737$& $0.723$& $0.564$& $0.509$& $0.502$& $0.501$& $0.5$ & $1/2$\\ \hline
	\end{tabular}
	\caption{[Model-I] Average of $r$ and $\mathrm{cos}\theta$ for different disorder strength $h$ using entire spectrum and the middle of the spectrum. The subscript $M$ in the first column represents the average extracted for the middle of the spectrum. Since model-I goes through a complex-real transition as the disorder is increased, the $\theta$ variable is ill-suited for $h\geq 6$. These parameter regimes are therefore represented by $\times$ in the row of $-\langle \mathrm{cos}\theta \rangle_{M}$.}
	\label{101}
\end{table}
\begin{table}[h!]
	\begin{tabular}{| l | l | l | l | l | l | l | l | l | l |}
	\hline
	&\multirow{2}{*}{$\mathrm{AI}^\dagger$}&\multicolumn{7}{|c|}{h} & \multirow{2}{*}{2D P} \\ \cline{3-9}
	&&$2$&$4$&$6$&$8$&$10$&$12$&$14$& \\ \hline
			$-\langle \mathrm{cos} \theta \rangle$ & $0.193$ & $0.16$& $0.039$& $-0.005$& $-0.011$& $-0.015$& $-0.021$& $-0.028$ & $0$ \\ \hline
	$\langle r \rangle$ &$0.722$& $0.722$& $0.686$& $0.669$&$0.668$& $0.667$& $0.667$& $0.666$& $2/3$ \\ \hline
	$-\langle \mathrm{cos} \theta \rangle_M$ &  $0.193$ & $0.186$& $0.071$& $0.007$& $-0.001$& $-0.003$& $-0.007$& $-0.014$ & $0$\\ \hline
	$\langle r \rangle_M$& $0.722$ & $0.722$& $0.692$& $0.669$& $0.667$& $0.667$& $0.667$& $0.667$ &  $2/3$ \\ \hline
	\end{tabular}
	\caption{[Model-II] Average of $r$ and $\mathrm{cos}\theta$  for different disorder strength using entire spectrum and middle of the spectrum (denoted by subscript $M$) at system size $L=18$.}
	\label{102}
\end{table}
\end{center}
\end{widetext}
\section{Complex level spacing ratio for middle of the spectrum}\label{appendix-e}
We recap that, in the main text, the marginal distributions of  $r$ and $\theta$ for the complex spacing ratio (CSR) were computed using the entire spectrum. These distributions in Figs.~\ref{fig:112}, \ref{fig:121}, \ref{fig:124} show good agreement with the marginal distributions for the random matrices and Poisson statistics in both the chaotic and the localized regimes respectively. However, there are minor discrepancies which are rooted in contribution coming from complex eigenvalues close to the edges. A common practice is to consider the middle of the spectrum. The prescription we employ when we refer to the middle of the spectrum is as follows. We consider $\pm 10 \%$ span of the complex spectrum from its centre/middle.
\begin{figure}[t]
\centering
\includegraphics[width=0.48\textwidth]{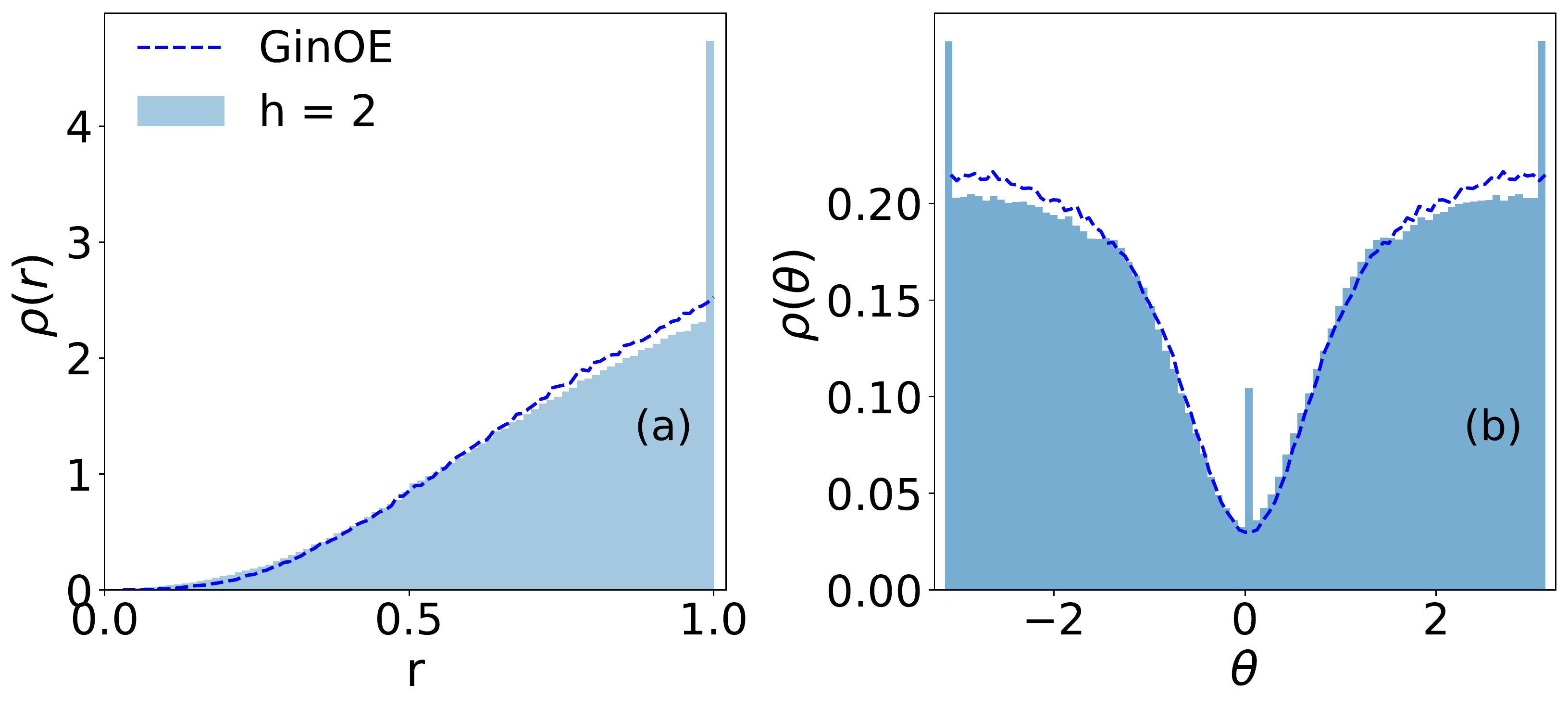}
\caption{[Model-I, Eq.~\ref{eq:model1}] Marginal distributions of (a) $r$ and (b) $\theta$ at $h=2$ for system size $L=18$ considering only middle of spectrum. Both the distributions agree with that of GinOE (blue Dashed).}\label{fig:11}
\end{figure}
\begin{figure}
\centering\includegraphics[width=0.48\textwidth]{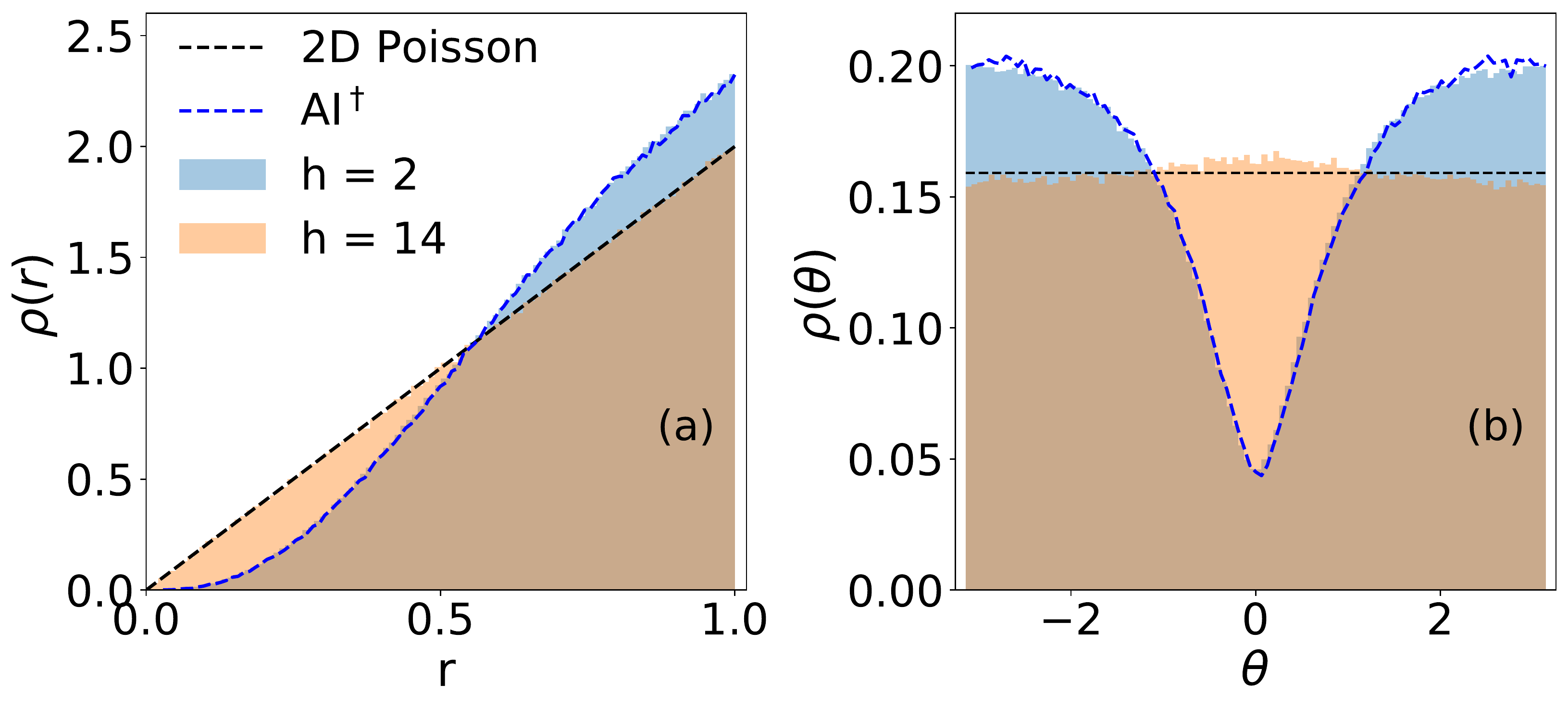}
\caption{[Model-II, Eq.~\ref{eq:model2}] Marginal distributions of (a) $r$ and (b) $\theta$ at $h=2$ (blue) and $h=14$ (orange) for system size $L=18$ considering only middle of the spectrum. The distributions agree perfectly with that of $\mathrm{AI}^\dagger$ symmetry class (blue dashed) and 2D Poisson statistics (black dashed) at $h=2$ and $h=14$ respectively.}
\label{fig:19}
\end{figure}
\begin{figure}
\centering\includegraphics[width=0.48\textwidth]{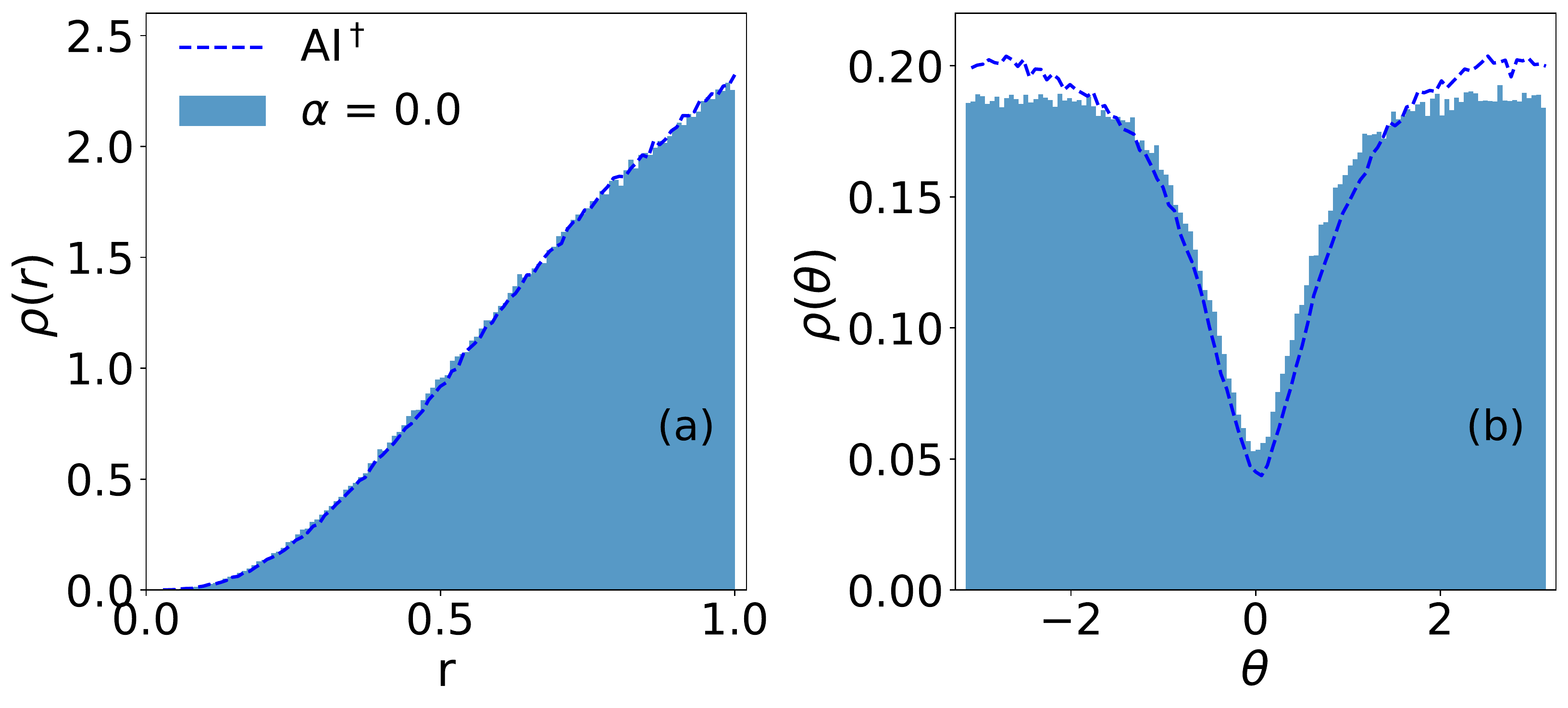}
\caption{[Model-III, Eq.~\ref{eq:model3}] Marginal distributions of (a) $r$ and (b) $\theta$ at $h=14$, $\alpha=0$ for system size $L=16$ considering only middle of the spectrum. The distributions agree well with that of $\mathrm{AI}^\dagger$ symmetry class.}
\label{fig:app-rho-3}
\end{figure}
\begin{table}
\centering
	\begin{tabular}{| l | l | l | l | l | l | l |}
	\hline
	&\multirow{2}{*}{$\mathrm{AI}^\dagger$}&\multicolumn{4}{|c|}{$\alpha$} & \multirow{2}{*}{2D P} \\ \cline{3-6}
	&&$0.0$&$0.5$&$1.5$&$4.0$& \\ \hline
	$-\langle \mathrm{cos} \theta \rangle$ & $0.193$ & $0.069$ & $-0.041$ & $-0.032$ & $-0.047$& $0$ \\ \hline
	$\langle r \rangle$ & $0.722$ & $0.715$ &  $0.7$ & $0.667$ & $0.664$ & $2/3$ \\ \hline	
	$-\langle \mathrm{cos} \theta \rangle_M$ & $0.193$ & $0.147$ & $0.104$& $-0.004$ & $-0.021$ & $0$ \\ \hline
	$\langle r \rangle_M$ & $0.722$ & $0.72$ & $0.705$& $0.67$ & $0.666$ & $2/3$ \\ \hline
	\end{tabular}
	\caption{[Model-III] Average of $r$ and $\mathrm{cos}\theta$ for different $\alpha$ at disorder strength $h=14$ using entire spectrum and middle of the spectrum (denoted by subscript $M$) at system size $L=16$.}
	\label{103}
	\end{table}
\begin{figure}[h!]
    \centering
    \includegraphics[width=0.34\textwidth]{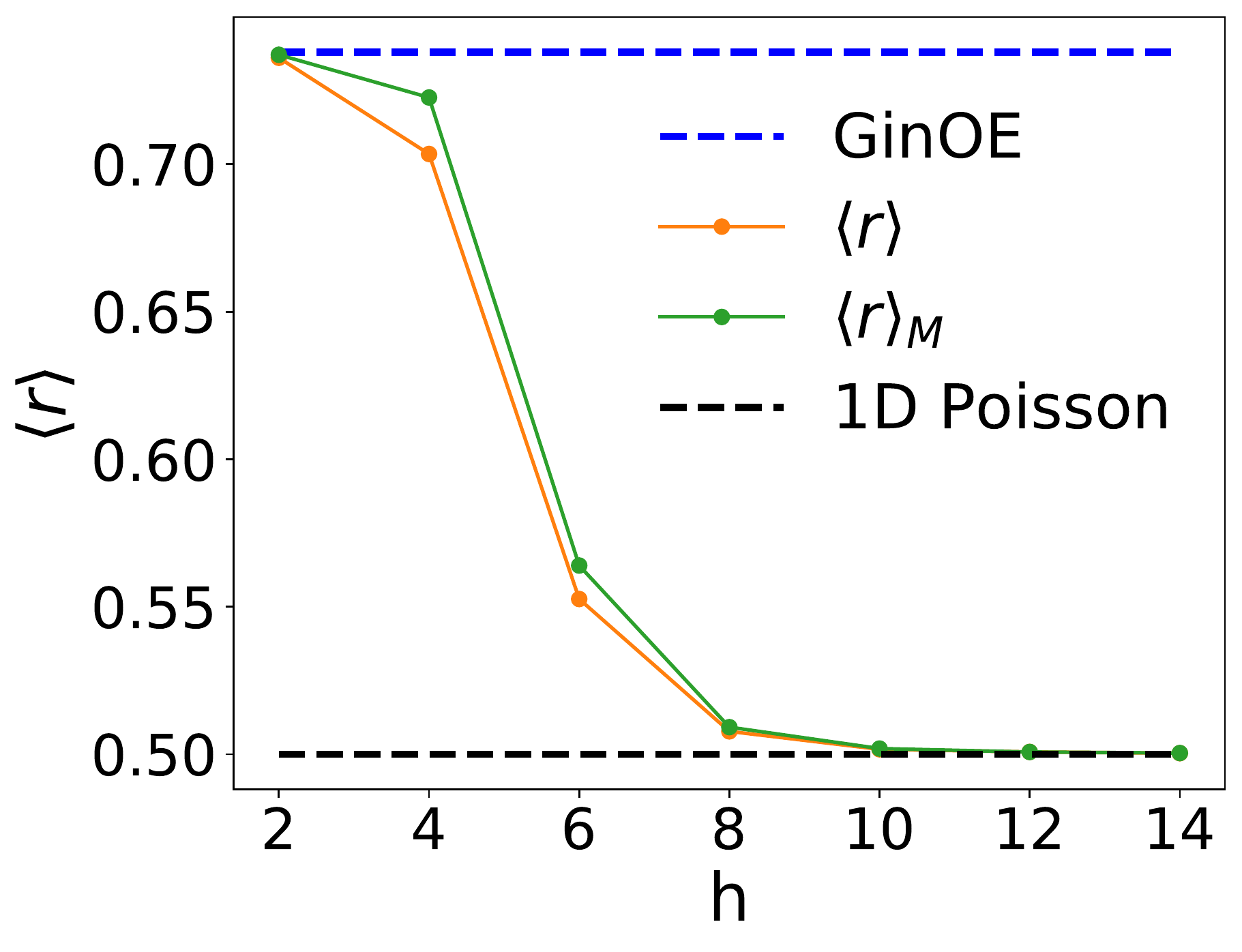}
    \caption{[Model-I, Eq.~\ref{eq:model1}] Average of $\langle r\rangle$ and $\langle r \rangle_M$ as a function of disorder strength $h$ for system size $L=18$. Here $\langle r \rangle_M$ is defined by taking only the middle of the spectrum.}
    \label{fig:raverage-I}
\end{figure}
\begin{figure}[h!]
    \centering
    \includegraphics[width=0.48\textwidth]{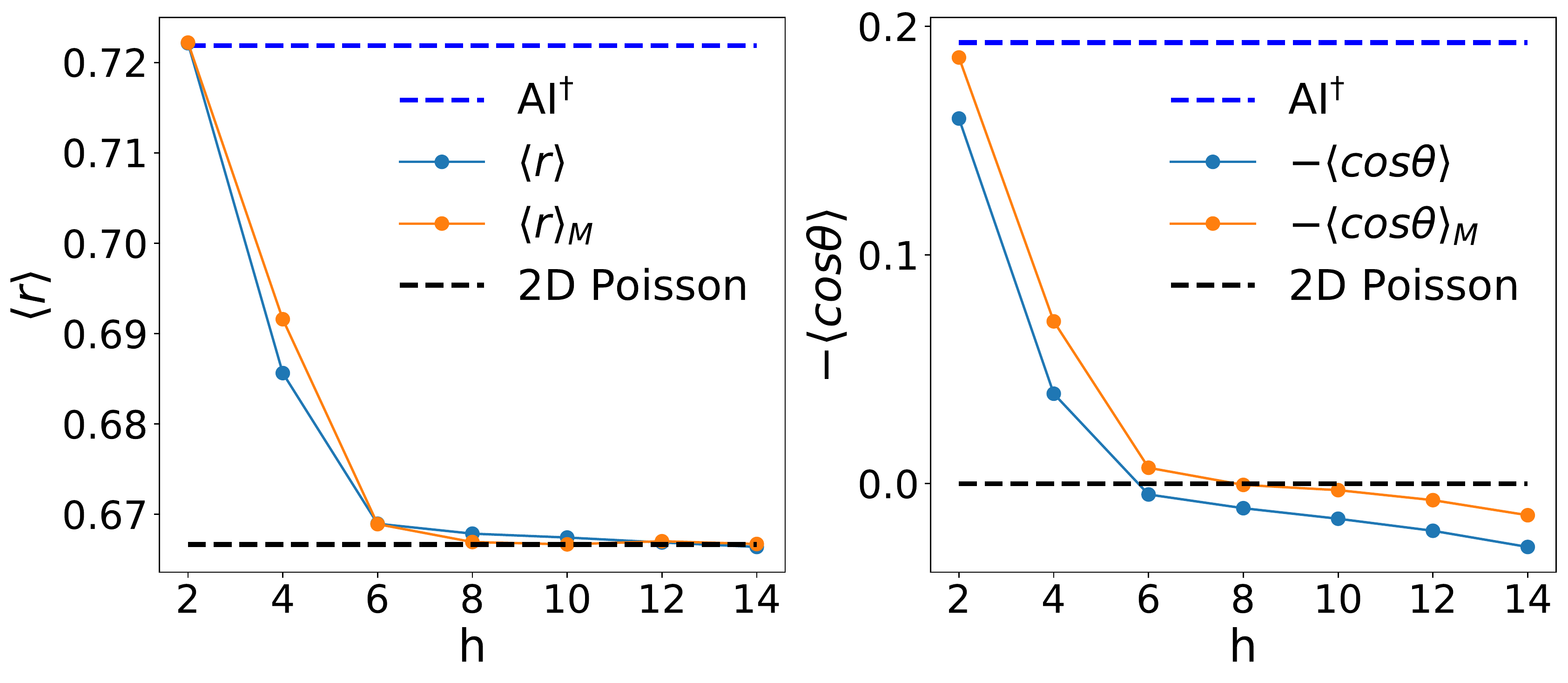}
    \caption{[Model-II, Eq.~\ref{eq:model2}] (a) Average of $\langle r\rangle$ and $\langle r \rangle_M$, and (b) average of $-\langle \mathrm{cos} \theta \rangle$ and $-\langle \mathrm{cos} \theta \rangle_M$ as a function of disorder strength $h$ for system size $L=18$. Here again, $\langle . \rangle_M$ is defined by taking only the middle of the spectrum.}
    \label{fig:raverage-II}
\end{figure}

In Fig.~\ref{fig:11}, we show the marginal distributions of $r$ and $\theta$ in the chaotic regime ($h=2$) for model-I (Eq.~\ref{eq:model1}). Both these distributions agree very well with GinOE and the comparison is much better than Fig.~\ref{fig:112} where the entire spectrum was considered. 
Fig.~\ref{fig:19} shows the spacing ratio distributions for $h = 2$ and $h=14$ with $L = 18$ for model-II (Eq.~\ref{eq:model2}). 
We can see that the marginal distributions are in perfect agreement with that of  $\mathrm{AI}^\dagger$ and 2D Poisson statistics at $h=2$ and $h=14$ respectively. The agreement is better than the one shown in Fig.~\ref{fig:121} which was made considering the entire spectrum. In Fig.~\ref{fig:app-rho-3}, we show similar agreement for the middle of the spectrum for model-III (Eq.~\ref{eq:model3}).

In addition to the marginal distributions, we also compute the average value of $r$ and $\mathrm{cos}\theta$ for all three models. These are tabulated in Tables  \ref{101},  \ref{102} and \ref{103}. The quantitative differences between the full spectrum and the middle of the spectrum, are also highlighted in the three tables. In Fig.~\ref{fig:raverage-I}, we show $\langle r\rangle $ versus disorder strength for model-I using both the entire spectrum and the middle of the spectrum. Similarly, in Fig.~\ref{fig:raverage-II}, shows $\langle r \rangle$  and $-\langle \mathrm{cos}\theta \rangle$ as a function of the disorder strength for model-II. 

%
\end{document}